\documentclass[aps,preprint,superscriptaddress,nofootinbib,preprintnumbers]{revtex4-1}
\usepackage{bm, amsmath,amssymb,amsfonts,slashed,array,graphicx,soul,subfigure,placeins}
\usepackage{multirow}
\usepackage{lineno}

\DeclareMathAlphabet{\mathpzc}{OT1}{pzc}{m}{it}

\newcommand{\bomb}{S}
\newcommand{\etav}{{\phi}}
\newcommand{\MET}{{E_T^{\text{miss}}}}
\newcommand{\mpstate}{\mathcal{B}}

\usepackage[dvipsnames]{xcolor}
\definecolor{red}{rgb}{0.9, 0,0}

\usepackage{hyperref}	
\graphicspath{{./Figures/}}

\makeatletter
\g@addto@macro\bfseries{\boldmath}
\makeatother

\begin{document}
\title{Triggering Soft Bombs at the LHC}

\author{Simon Knapen}
\affiliation{Ernest Orlando Lawrence Berkeley National Laboratory, University of California, Berkeley, CA 94720, USA}
\affiliation{Department of Physics, University of California, Berkeley, CA 94720, USA}
\author{Simone Pagan Griso}
\affiliation{Ernest Orlando Lawrence Berkeley National Laboratory, University of California, Berkeley, CA 94720, USA}
\author{Michele Papucci}
\affiliation{Ernest Orlando Lawrence Berkeley National Laboratory, University of California, Berkeley, CA 94720, USA}
\affiliation{Department of Physics, University of California, Berkeley, CA 94720, USA}
\author{Dean J. Robinson}
\affiliation{Ernest Orlando Lawrence Berkeley National Laboratory, University of California, Berkeley, CA 94720, USA}
\affiliation{Department of Physics, University of California, Berkeley, CA 94720, USA}
\affiliation{Physics Department, University of Cincinnati, Cincinnati OH 45221, USA}

\begin{abstract}
Very high multiplicity, spherically-symmetric distributions of soft particles, with $p_T \sim \text{few}\times 100$~MeV, may be a signature of strongly-coupled hidden valleys that exhibit long,
efficient showering windows. With traditional triggers, such `soft bomb' events closely resemble pile-up and are therefore only recorded with minimum bias triggers at a very low efficiency. 
We demonstrate a proof-of-concept for a high-level triggering strategy that efficiently separates soft bombs from pile-up by searching for a `belt of fire': A high density band
of hits on the innermost layer of the tracker. 
Seeding our proposed high-level trigger with existing jet, missing transverse energy or lepton hardware-level triggers, we show that net trigger efficiencies of order $10\%$ are possible for
bombs of mass $\text{several}\times100$~GeV. We also consider the special case that soft bombs are the result of an exotic decay of the 125 GeV Higgs. The fiducial rate for `Higgs bombs' triggered in this manner is marginally higher than the rate achievable by triggering directly on a hard muon from associated Higgs production.

\end{abstract}

\maketitle

\tableofcontents

\clearpage

\section{Introduction}
In the  last five to ten years, our understanding of the possible hidden extensions of the Standard Model has expanded dramatically. These new examples yield a wide range of spectacular but often subtle signatures at LHC, which more often than not require very refined search strategies.
At the same time, in the absence of compelling signs of more conventional beyond the standard model physics in the LHC Run~I and in the first part of Run~II, innovative trigger strategies for ATLAS, CMS and LCHb are becoming an important frontier. A particularly interesting experimental development on this front is the capability to  carry out certain searches (partially) \emph{online}, a technique often referred to as `data-scouting'  or `trigger-level data analysis'. It has proven to be a particularly effective tool in extending the sensitivity of ATLAS and CMS to dijet resonances down to \mbox{$\sim$ 500 GeV}~\cite{Khachatryan:2016ecr,ATLAS-CONF-2016-030}.  Such strategies must necessarily be implemented \emph{before} the bulk of the data is collected, and as such their 
development is especially urgent now that LHC is beginning to transition to its luminosity-driven phase. It is therefore presently incumbent on phenomenologists to explore new ways to leverage (partially) online searches, in order to search for well-motivated, though perhaps unexpected, new physics signals. See \cite{Ilten:2015hya,Ilten:2016tkc} for examples of efforts in this direction in the context of LHCb.

In this paper we propose such a partially online strategy for ATLAS and CMS, in which one searches for events that are primarily characterized by an anomalously large particle multiplicity: 
A case that has received comparatively little attention. This type of event may be a signal of a strongly coupled hidden valley~\cite{Strassler:2006im}. Hidden valleys (HV) with 
confining dynamics are themselves well-motivated extensions of the standard model (SM), that provide generic 
ingredients for use in for instance models of dark matter~\cite{Kribs:2009fy} or neutral naturalness~\cite{Chacko:2005pe,Burdman:2006tz}. (See also~\cite{Kribs:2016cew} for a recent review of the role of hidden strong dynamics in dark matter models.) The characteristic feature of a hidden valley is that they can be accessed efficiently via a heavy portal 
state, whose mass is much larger than the mass gap of the confining dynamics. All energy deposited into the hidden valley by a collider is then distributed over the states at the bottom 
of the hidden valley spectrum, generating a comparatively higher multiplicity of final states. These states may be dark pion~\cite{Strassler:2006im}, glueball~\cite{Juknevich:2009gg,Juknevich:2009ji} or onium states~\cite{Han:2007ae}, some of which may decay promptly or displaced back to the SM sector.

Significant progress has been made in the recent years in developing searches for displaced 
decays, to the extent that both ATLAS and CMS now have a set of dedicated  trigger strategies~\cite{Aad:2013txa,Perrotta:2015jyu}, vertex finding algorithms~\cite{Aad:2013ela} 
and various analyses for lepton-jets~\cite{CMS:2015pca,CMS:2014hka,Chatrchyan:2012cg,Chatrchyan:2011hr,Aad:2012kw,Aad:2012qua,Aad:2015sva,Aad:2014yea,Aad:2015sms,ATLAS:2016jza}, displaced 
decays~\cite{CMS:2014wda,CMS:2015sjc,CMS:2016isf,Chatrchyan:2012ir,Chatrchyan:2012jwg,CMS:2015gga,ATLAS:2012av,Aad:2015asa,Aad:2015uaa,ATLAS:2016olj,Aad:2015rba,Aad:2012zx,Aad:2011zb,TheATLAScollaboration:2013yia} 
and unparticles~\cite{Khachatryan:2015bbl,CMS:2016yfc,Khachatryan:2014rra} that have been performed. In 
addition to these ongoing efforts, new specialized off-line strategies have been proposed for, \emph{e.g.},~semi-visible jets~\cite{Cohen:2015toa} and emerging jets~\cite{Schwaller:2015gea}. 
Still, various signatures arising in HV phenomenology remain without concrete search strategies that can be implemented with efficient triggers and suitable offline analyses, \emph{e.g.}~\cite{Kang:2008ea,Harnik:2008ax}. 
Here we instead focus 
on one of these cases: A strongly coupled, quasi-conformal hidden valley, whose content promptly decays to the visible sector, and whose mass gap is much smaller than the mass of the portal state~\cite{Strassler:2008bv}. 
(The fully invisible case corresponds to the unparticle scenario~\cite{Georgi:2007ek}.) 

A long showering window combined with a 
large `t~Hooft coupling results in efficient showering into a nearly spherically-symmetric, soft spray of particles at the bottom of the hidden valley spectrum. If these particles can decay promptly back to the SM sector, a so-called `\emph{firework}' or `\emph{soft bomb}' is created in the detector. In this work we consider soft bombs originating from two different production mechanisms: soft bombs generated by the decay of a heavy exotic spin-0 state, produced through gluon fusion (GF);  and soft bombs generated by the decay of the 125 GeV Higgs -- a `Higgs bomb' -- which may be generated by gluon fusion, 
vector boson fusion (VBF), or associated production (VH). Soft bombs generated by decays of heavy vectors, produced, \emph{e.g.}, via $q\bar{q}$ fusion, will have similar phenomenology to the scalar 
soft bombs explored in this work, and similar conclusions should apply.

The sphericity of the soft bomb event, as well as its lack of hard, isolated objects, implies that the signal strongly resembles pile-up, especially at the calorimeter level. 
As a result it would generally not be picked up by conventional triggers. Na\"\i vely, it would seem possible to select a soft bomb event by counting the number of tracks corresponding 
to the primary vertex. While information from the inner detector is available to the high level trigger (HLT), 
the expected softness of the soft bomb $p_{\rm T}$ spectrum, together with the computing requirements needed to run track reconstruction at HLT, makes
this approach far from plausible. 

We propose instead to bypass the slow step of track reconstruction and directly trigger on a \emph{`belt of fire'}, meaning a ring-shaped overdensity of hits
on the inner layers of the tracking system.  Specifically, our proposed triggering strategy is as follows:
\begin{enumerate}
	\item Hard jets from initial state radiation permit a moderate fraction of events to pass the existing level 1 (L1) trigger. (For VBF and VH production of Higgs bombs, associated hard jets 
	or leptons permit the same.) Moreover, a sizable fraction of the final states -- so called `loopers' -- are too soft to reach the electromagnetic calorimeter (ECAL), as shown schematically 
	in Fig.~\ref{fig:SMET}. This means that a soft bomb recoiling against a hard object can generate sizable $\MET$, and thereby also pass the (L1) $\MET$ trigger with a reasonable efficiency.
\begin{figure}[t]
	\includegraphics[width=6cm]{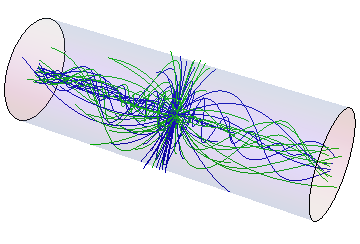}
	\caption{Schematic representation of a soft bomb event with $\sim 100$ tracks, showing electrons and muons in blue and 
green respectively. The cylinder represents the inner boundary of the ECAL. An $\mathcal{O}(1)$ fraction of the tracks are too soft to reach the 
 ECAL, generating $\MET$ if the bomb itself is recoiling against other hard particles in the event.}
	\label{fig:SMET}
\end{figure}

	\item At the HLT level, we search for a highly localized population of hits compared to the more diffuse background from pile-up interactions. To minimize the spreading of the signal hits, 
	we focus on the innermost layer of the tracker.

	\item In an off-line analysis it should be possible to fully reconstruct the event, and enhance background rejection via requirements on
	track multiplicities.  In addition, it may be possible to extract extra information from the factorial moments and cumulants of the multiplicity distributions \cite{Sanchis-Lozano:2015eca}. Variables based on the track multiplicity are also promising for more weakly coupled hidden valleys  \cite{Strassler:2008fv}.

\end{enumerate}
To explore the efficacy of this strategy, we simulate soft bomb generation and propagation inside a simplified model of the 
ATLAS detector for a number of representative benchmark points and estimate the signal efficiencies that can be obtained at 
both stages of the trigger.  We show that the triggering efficiencies for bombs of mass $\text{several}\times100$~GeV could be as high as $\sim10\%$. 
Further, the acceptance rate for Higgs bombs triggered in the manner is comparable to the rate achievable by triggering directly on a hard muon in the associated production.

This paper is organized as follows. In Sec.~\ref{sec:models} we briefly review the features of strongly coupled hidden 
valleys, describe possible portals to the SM sector and present the benchmark models studied in the later sections. Section~\ref{sec:strategy} contains our 
proposed trigger strategy and corresponding efficiency estimates.
We conclude in Sec.~\ref{sec:conclusions} and reserve details regarding our simulations and validation to Appendix~\ref{app:detecsim}.

\section{Soft bomb framework}\label{sec:models}

A soft bomb event is generically represented by the process $pp \to \mpstate +X$, where $\mpstate$ is a multi-particle state of soft SM particles with very large multiplicity -- $N \sim 10^2$ to $ 10^4$ -- roughly spherically distributed in the center-of-mass frame of $\mpstate$~\cite{Strassler:2008bv}. Such events may be generated by portals between the SM and a confining hidden valley, with appropriate fragmentation features and hadronization behavior. 

Soft bomb events may be generated via a fairly broad variety of portals, mostly differing by mediation mechanisms from and to the visible sector and their corresponding signatures. In selecting representative portals, discussed below, we are guided not by theoretical considerations such as minimality or simplicity of the model, but by experimental ones, choosing portals that represent the worst-case scenarios for current triggers. We reserve a more thorough investigation of the various experimental signatures, including different final states and non-prompt decays, to future work.

\subsection{Fragmentation}
In a non-abelian gauge theory in the perturbative regime, radiation is unsuppressed only in the collinear or soft regions of phase space, which are enhanced
by the presence of large logarithms. However, as the ('t~Hooft) coupling of the gauge theory becomes larger and larger, one expects the large angle emission of partons 
carrying an $\mathcal{O}(1)$ fraction of the momentum to be correspondingly more and more significant. This leads to a more isotropic distribution of partons, without
large hierarchies in energy between them. This behavior is observed in, \emph{e.g.}, low energy hadronic interactions, although in this case the overall parton multiplicity is low because of the small energy range in which $\alpha_s(Q^2)$ is large, where here and hereafter $Q$ denotes the scale of the hard process. Since a high degree of sphericity (in its center-of-mass frame) is one of the requirements for $\mpstate$, we are naturally led to consider hidden valleys that are strongly coupled over a sizable energy range. 
 
Regardless of the strength of the `t~Hooft coupling, the parton multiplicity generated during the evolution of the  shower is controlled by the ratio $Q/\Lambda$, where $\Lambda$ is the hadronization scale. In particular, using parton-hadron duality, one can relate the average hadron multiplicity  
to the average number of partons obtained by evolving the system from $Q$ down to $\Lambda$. The latter is the first ($j=1$) Mellin moment of the fragmentation function~\cite{Ellis:1991qj}. At sufficiently low parton energy fractions, $1-x \gg 1/\sqrt{\lambda}$, with $\lambda$ the 't~Hooft coupling, 
it is controlled by the timelike (fragmentation) anomalous dimension, $\gamma_T(j)$ evaluated at $j=1$. 
For example in QCD-like theories in the perturbative regime, one obtains~\cite{Ellis:1991qj}
\begin{equation}
\langle n(Q)\rangle \propto \exp\left[\frac{1}{b}\sqrt{\frac{6}{\pi \alpha_s(Q^2)}}+\left(\frac{1}{4}+\frac{5 n_f}{54 \pi b}\right)\log \alpha_s(Q^2)\right]
\end{equation}
which is valid for sufficiently large values of the first coefficient of the $\beta$ function, $b$.
In the case where the running of the gauge coupling can be neglected (which is going to be the relevant case in the following), the average parton multiplicity reduces to
\begin{equation}
\langle n(Q)\rangle \propto \left(\frac{Q}{\Lambda}\right)^{2\gamma_T(j)}\bigg|_{j = 1}\,.
\end{equation}
Since $\gamma_{T}(1)>0$, it is clear from this expression that the $Q/\Lambda$ ratio should be sizable to obtain a large number of partons. 

This observation implies that the 't~Hooft coupling should remain strong over a large energy window without triggering the generation of a mass gap. That is, the HV must contain a {\it walking} or quasi-conformal, strongly-coupled hidden sector below the SM--HV mediation scale. In such theories, AdS/CFT or conformal symmetry arguments become available, and may provide useful information regarding the behavior of the HV and the dependence of $\gamma_{T}(j)$ on $\lambda$. 
Moreover, if one applies the Gribov-Lipatov (GL) relation~\cite{Gribov:1972rt} between the spacelike and timelike anomalous dimensions, one can determine the energy dependence of $\langle n(Q)\rangle$ as a function
of the 't~Hooft coupling~\cite{Hatta:2008tn}. This relation has been studied for $N=4$ SYM at strong coupling via AdS/CFT~\cite{Basso:2006nk,Gubser:2002tv} as well as perturbatively up to three loops in various gauge theories~\cite{Dokshitzer:2005bf,Dokshitzer:2006nm}. In the perturbative studies, its breakdown has been found to be proportional to the first $\beta$ function coefficient and hence to vanish for a CFT.  We will therefore assume the approximate validity of the GL
relation for our HV sector, such that one finds~\cite{Hatta:2008tn}
\begin{equation}
	\label{eq:multiplscaling}
	\langle n(Q) \rangle \propto \left(\frac{Q}{\Lambda}\right)^{1 +\mathcal{O}(1/ \sqrt{\lambda})}\,,
\end{equation}
for $\lambda \gg 1$. This is consistent with a picture of unsuppressed emission of a large number partons, all sharing a similar amount of energy. In the large $\lambda$ limit it is moreover possible to compute the two-point function of the stress-energy tensor in the weakly coupled AdS dual, and to show that to leading order the soft bomb events ought to be spherically symmetric in the center-of-mass frame~\cite{Polchinski:2002jw,Hofman:2008ar,Lin:2007fa}. 

Equation \eqref{eq:multiplscaling} is suggestive of a statistical ensemble, and therefore also provides a sense of the distribution of hadronic energy and momentum. Studies of models of meson multiplicities, beginning with the Fermi statistical model~\cite{Fermi:1950jd}, and followed by the Hagedorn fireball picture~\cite{Hagedorn:1965st}, 
the Bjorken-Brodsky model~\cite{Bjorken:1969wi} and more recently phenomenological fits to QCD data and AdS/CFT calculations~\cite{Hatta:2008qx}, all point to a description in which the high energy tail is exponentially suppressed, 
approximately following a thermal (either Maxwell-Boltzmann~\cite{Becattini:2001fg} or Tsallis~\cite{Cleymans:2012wm}) distribution characterized by a ``temperature'' of the order of the confinement scale. In particular in gauge theories with gravity duals, it has been
found that $T/\Lambda \sim 1$-$ 2.5$~\cite{Hatta:2008qx}, while pion and kaon spectra in hadronic collisions are well fitted by $T \sim 160$-$190\,{\rm MeV}$~\cite{Becattini:2008tx,Becattini:2010sk,Becattini:2009ee,Ferroni:2011fh}.

In this work, since we are mostly interested in the leading order characteristics of these events, we will assume a simplified picture of the fragmentation, in which mesons are spherically distributed, with a Maxwell-Boltzmann momentum distribution
\begin{equation}\label{eq:thermal}
	\frac{d N}{d^3{\bf p}} \sim \exp\Big\{- \sqrt{\bm{p}^2 + m^2}/T\Big\}\,.
\end{equation}
Here the temperature $T$ and meson mass $m$ are both of order $\Lambda$. (We elaborate on our assumptions regarding the meson spectrum in the next section.) Should soft bomb events be observed, studying the deviations from these assumptions would provide valuable information on the HV gauge sector.

\subsection{Hadronization}
Near the hadronization scale $\Lambda$ one typically expects a rich spectrum of bound states. The detailed spectrum of mesons and baryons is highly dependent upon the HV degrees of freedom, but generically 
encompasses states with masses of $\mathcal{O}(\Lambda)$ and parametrically lighter pNGB ``pions'' if the mass-generation mechanism breaks some of the global symmetries of the theory. Therefore one expects states
with $m \gtrsim T$ and potentially with $m \lesssim T$. The final energy spectra of the fragmentation products depend both on the energy distributions described in the previous subsection (Eq.~\eqref{eq:thermal}) and on the hadron decay
chains. In particular, prompt decay of the heavier hadronic states may introduce sub-leading non-thermal populations of lighter, daughter hadrons, further broadening the prompt thermal distributions assumed in the fragmentation.
Moreover, the presence of (approximate) unbroken residual global symmetries in the hidden sector may render some of the hadrons (meta-)stable. 

In order to capture the leading features of these hadronization models,  in this work we make the simplifying assumption that the low lying spectrum is modeled by a single flavor of a light (pseudo)scalar meson, $\etav$, that decays promptly to soft SM states, with $m_{\etav} \sim T$. This assumption likely corresponds to a worst-case scenario as far as conventional triggering strategies are concerned. For instance, the presence of neutral, stable states will slightly worsen the efficiency of our tracker-based, HLT strategy, but at the same time will be a source of additional $\MET$. This extra $\MET$ enhances the L1 trigger efficiency, which we expect to more than offset a possible loss at the HLT, as long as the majority of the hidden sector states still decay promptly. Second, the presence of states significantly lighter than $T$ should lead to more energetic decay products and would increase the likelihood of passing some of the triggers that we discuss below. We therefore expect that the effects arising from more complicated spectra do not significantly deteriorate our analysis and conclusions.  

For our benchmark study, we choose $T$ somewhat smaller than $m_{\etav}$. This softens the spectrum in \eqref{eq:thermal}, which is an attempt to conservatively account for some of the theory uncertainty. In order to maximize softness we always require the daughter SM particles of the $\etav$ decay to be close to the detectable thresholds at ATLAS or CMS, that is, with energies $\sim \text{few}\times100$~MeV, and adjust relevant thresholds accordingly. Anticipating $4$-body SM final states for $\etav$ decays, throughout our study we choose as a benchmark
\begin{equation}
	\label{eqn:MTB}
	m_{\etav} =1\,\text{GeV}\,,\qquad  \text{and} \qquad  T= 0.5\,\text{GeV}\,.
\end{equation}

\subsection{Production}
The mediation between the SM and the HV can be achieved by coupling a massive state in the visible sector to some operator of the HV. Schematically
\begin{equation}\label{eq:portal}
	\mathcal{L} \supset   M^{4-\Delta_{\mathrm{vis}}-\Delta_{HV}}\mathcal{O}_{\mathrm{vis}} \mathcal{O}_{HV}\,,
\end{equation}
with $\mathcal{O}_{\mathrm{vis}}$ and $\mathcal{O}_{HV}$ operators in the visible sector and in the HV respectively. $\Delta_{\mathrm{vis}}$ and $\Delta_{HV}$ are the corresponding operator dimensions and $M$ the mass scale associated with the portal. In the terms of the parton level degrees of freedom, $\mathcal{O}_{HV}$ may be for instance $G'_{\mu\nu}G'^{\mu\nu}$ or $\psi'\bar\psi'$, which represent the hidden sector gauge fields and fermions respectively. From an IR point of view, $\mathcal{O}_{HV}$ can be seen as a linear combination of terms of the form $\partial^{2P}\phi^N$, where in our models $N\gg1$. The operator $\mathcal{O}_{\mathrm{vis}}$ is part of the visible sector, and could consist of SM fields or exotic heavy mediator fields.
In either case, $\mathcal{O}_{\mathrm{vis}}$ should be constructed such that HV states can be produced by quark or gluon collisions at an appreciable rate at LHC. The HV phenomenology we focus on in this paper generally requires that $\mathcal{O}_{\mathrm{vis}}$ is color and electrically neutral and that the operator in Eq.~\eqref{eq:portal} does not spoil the quasi-conformal nature of the hidden sector. Partially relaxing these assumptions may lead towards certain types of
quirky phenomenology~\cite{Kang:2008ea}. While some of the ideas proposed in this work may be applied to those cases, a systematic study is beyond the scope of this paper. The states associated with $\mathcal{O}_{\mathrm{vis}}$ can be
either singly or pair produced, the latter as in the case of, \emph{e.g.}, a fermionic operator. In the latter case each event then contains two soft bombs. 

\begin{figure}[t]
	\includegraphics[width=7cm]{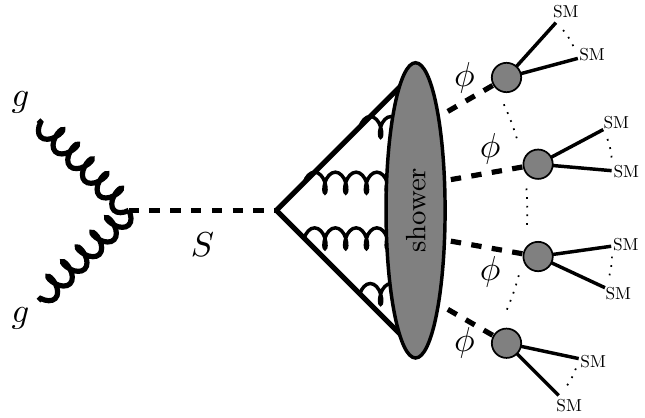}
	\caption{Schematic gluon fusion production of the mediator, $S$, followed by quasi-conformal showering and $\etav$ decay, generating a heavy soft 
bomb event. If the Higgs is the mediator, production may also occur through vector boson fusion or with an associated vector boson, and with 
identical showering and decay.}
	\label{fig:HSBS}
\end{figure}

For concreteness, we now focus on the case of single resonant production 
of a massive (pseudo)scalar mediator $\mathcal{O}_{vis}=\bomb$, and leave a more systematic exploration of non-resonant or pair production cases to future work.  We hereafter assume gluon fusion (GF) production via the dimension-five 
operators $\bomb G_{\mu\nu} G^{\mu\nu}$ or $\bomb G_{\mu\nu} \tilde{G}^{\mu\nu}$.  Production and subsequent showering into the 
dark HV can then be represented diagrammatically as shown in Fig.~\ref{fig:HSBS}.  The UV completion of 
this portal likely involves exotic heavy colored states, which we do not consider here. An alternative production mode may be via a heavy vector which couples to the SM quarks. Up to deviations induced by the difference between the gluon and quark parton distribution functions, we expect the phenomenology of such a vector-portal to be qualitatively similar to the scalar-portal case, so we consider only the former.

A second type of benchmark we consider is the case where the soft bomb is produced by an exotic decay of the SM-like Higgs scalar, and can therefore be produced through gluon fusion (GF) vector boson fusion (VBF) or with an associated vector boson (VH). The simplest possibility to generate this decay mode is through an operator of the form $ H^\dagger H\mathcal{O}_{HV}$. However in this setup the Higgs vacuum expectation value will introduce a tadpole for $\mathcal{O}_{HV} $, which generically breaks conformal invariance not too far from the weak scale, see \emph{e.g.}~\cite{Fox:2007sy,Delgado:2007dx,Kikuchi:2007qd}. This may introduce a tension with any approximate conformality in the HV sector below the EW scale, as assumed above. Some fine-tuning or more detailed model building may therefore be needed to postpone the breaking of conformal symmetry to the GeV-scale, while maintaining an appreciable branching ratio to the hidden sector. For instance, the hidden sector could be made to flow to new fixed conformal point well below the weak scale, preserving conformal dynamics over parts of the energy window between the weak scale and the scale of the light hadronic states~\cite{Nelson:2009jg}. This potential tadpole problem might also be bypassed entirely if $S$ couples to the SM only via quartic interactions, \emph{i.e.},  with couplings  of the form $S S^\dagger HH^\dagger + S\mathcal{O}_{HV}$ and $m_{S}<m_h/2$. In this case one must, however, consider soft bomb pair production.

\subsection{Decay channels}
\label{sec:DC}
Portals permitting HV states to decay back to the SM sector have been already widely discussed. Depending on the mechanism, both prompt and displaced decays are possible into pair or multi-body final states. In this work we focus on prompt $\etav$ decays, which we believe are the most challenging scenarios for high multiplicity final states from a trigger perspective. The case of displaced decays is deferred to future work.

Specifically, we assume $\etav$ is coupled to hidden photons via $\phi A'^{\mu\nu}A'_{\mu\nu}$ or $\phi A'^{\mu\nu}\tilde A'_{\mu\nu}$ operators, that are kinetically mixed with SM hypercharge via the operator $\varepsilon A'^{\mu\nu} B_{\mu\nu}/2$. Such dimension-five operators may arise by weakly gauging some hidden sector U(1) global symmetry, as long as the coupling is small enough that it does not spoil the quasi-conformal character of the HV. The two-body decay modes $\etav  \to \ell \bar{\ell}$ or $\etav \to \pi \pi$ cannot be generated at tree-level through this kinetic mixing portal. We assume then, that $m_{\etav} > 2m_{A'}$, so that $\phi$ may instead decay promptly to two on-shell $A'$'s.  
Various subsequent prompt $A'$ decay modes to SM degrees of freedom then may proceed through the kinetic mixing portal, depending 
on thresholds set by the $A'$ mass. A single $\etav$ then typically decays to at least four SM degrees of freedom and the $\etav$ decay 
to electrons is not chirally suppressed compared to muons and pions. 

Following from Eq.~\eqref{eqn:MTB}, we select as our benchmark
\begin{equation}
	 2 m_\mu<  m_{A'} < 500\,\text{MeV} \,,
\end{equation}
for which 
\begin{equation}
\Gamma[A' \to \ell^+ \ell^-] \simeq \frac{m_{A'}\alpha \varepsilon^2 \cos^2(\theta_W)}{3}  \left(1 +  \frac{2m_\ell^2}{m_{A'}^2}\right) \sqrt{1 - 
\frac{4m_\ell^2}{m_{A'}^2}}\,.
\end{equation} 
For \mbox{$m_{A'} \gtrsim 250$~MeV}, the decay rate corresponds to $c\tau \lesssim 1$\,mm for $\varepsilon \gtrsim 10^{-5}$. 
This leaves plenty of non-excluded dark photon $\varepsilon$--$m_{A'}$ parameter space~\cite{Alekhin:2015byh}, that generates a sufficiently prompt $A'$ 
decay. Compared to the dilepton rate, the exclusive decay $A' \to \pi^0\gamma$ is suppressed by $\alpha$ while $A' \to \pi^+\pi^-\pi^0$ is phase-space suppressed. (Alternatively, these suppressions can be understood by assuming vector meson dominance, so that these decays proceed dominantly via $A'$ - $\omega/\rho^0$ mixing.)

For the $A'$ mass range considered here, $\etav$ decays access final states composed of electrons, muons and pions with comparable abundances. Pions are expected to behave similarly to muons in our trigger analysis, except that they would deposit a larger amount of energy in the calorimeter. We therefore focus on final states composed only of electrons and muons, and assume that $A'$ decays to $e^+e^-$ and $\mu^+\mu^-$ with equal branching ratios. This allows us to study the effects of both tracks and bremsstrahlung photons (produced by the electron population) on the trigger efficiencies. We emphasize that these assumptions are made for simplicity only, and the general conclusions of our analysis should also apply for other choices regarding the $\etav$ decay modes, so long as an $\mathcal{O}(1)$ fraction of the decay products carry electric charge. 

\section{Trigger analysis}
\label{sec:strategy}
\subsection{Simulation}

For the quasi-conformal, strongly-coupled hidden valley models considered herein, the $p_T$ spectrum of the soft bomb particles is determined only by the meson mass $m_\etav$ and the temperature $T$ -- both of the order of the hadronization scale $\Lambda$ -- as in Eq.~\eqref{eq:thermal}. The efficient showering ensures that the multiplicity, $N$, of final state particles scales as
\begin{equation}
	N \sim m_S/m_\etav\,.
\end{equation}
Hence the mass of the mediator itself, $m_{\bomb,h}$, affects only the multiplicity of the particles in 
the final state, as well as the total energy deposited in the detector, but not the $p_T$ spectrum. (This differs from other, typical perturbative hidden valley models, \emph{e.g.}~\cite{Baumgart:2009tn}, in which the mass scale of the mediator sets the $p_T$ spectrum of the 
final states, while the multiplicity of the final states is roughly independent of the mediator mass.)

Our goal is to develop a versatile trigger strategy for soft bomb models, which is sensitive even to the most difficult cases. As in Eq.~\eqref{eqn:MTB}, for the assumed kinetic mixing portal of Sec.~\ref{sec:DC} we therefore have chosen both $m_\etav$ and $T$ such that the final states have $p_T \sim \text{few}\times100$ MeV, which is 
near the lower threshold of what ATLAS and CMS could plausibly observe. To characterize the increasing difficulty of observing the soft bomb as the mass of the mediator decreases, and hence 
as its particle multiplicity and total energy deposition decreases, we focus on three benchmark points: mediator masses of 750~GeV, 400~GeV, and 125~GeV, which correspond to a high, medium and low mass 
benchmark point, respectively. The choice of the $125$~GeV benchmark is motivated by the Higgs, which may itself serve a portal into the hidden sector.  For the high and medium mass benchmarks, 
we assume production through gluon fusion (GF), but our analysis is also applicable to topologies with a $q\bar q$ initial state via a vector portal. For the Higgs portal benchmark we consider production 
via gluon fusion, vector boson fusion (VBF) and associated production (VH), with correspondingly different trigger paths. Our benchmarks are summarized in Table~\ref{tab:sigbenchmark}.  

\begin{table}[t]
\renewcommand*{\arraystretch}{1.1}
\newcolumntype{C}{ >{\centering\arraybackslash} m{2cm} <{}}
\begin{tabular}{|C|C|C|c|}\hline
	&$m_{\etav}$~(GeV) &$T$ (GeV)&Production\\\hline\hline
	750 GeV&1.0&0.5&GF\\
	400 GeV&1.0&0.5&GF\\
	Higgs&1.0&0.5&GF, VBF, VH\\
	\hline
\end{tabular}
\caption{Signal benchmark points. \label{tab:sigbenchmark}}
\end{table}

For the gluon fusion benchmarks, we generate a suitable matched sample of $\bomb/h+0,1j$ with \texttt{MadGraph 5 v2.3.3}~\cite{Alwall:2014hca} and \texttt{Pythia 8.212}~\cite{Sjostrand:2006za,Sjostrand:2014zea} 
for 13~TeV proton-proton collisions, in which we treat $\bomb$ as a stable scalar boson. For the VBF and VH samples only \texttt{Pythia 8} is used, 
and no matching is performed. The stable scalar boson in these samples is then replaced with a soft 
bomb, which is generated with our own Monte Carlo code according to the requirements described in Sec.~\ref{sec:models}.
Since we shall stick with existing trigger paths for the L1 trigger, all relevant background rates have been measured already. For the HLT, since we propose an entirely new software-based trigger, 
we must therefore estimate the level of background rejection from simulation. Anticipating that jet and missing transverse energy ($\MET$) triggers will be the most efficient pathways 
to pass the L1 trigger for gluon fusion production, 
we generate both a dijet sample and a $Z+0,1j$ with $Z\rightarrow \nu\nu$ sample to estimate the background rejection rates for our HLT strategy. 
The background in our HLT strategy however turns out to be heavily dominated by the properties of pile-up, and in practice 
we find that the nature of the hard process itself is irrelevant for the HLT observables we will propose. For this reason the dijet sample was generated only to leading order with \texttt{Pythia 8}.

 For both signal and background, we include pile-up in the form of minimum bias events generated with \mbox{\texttt{Pythia 8}}, where we assume the number of pile-up vertices follows a Poisson 
 distribution with an average of 50 interactions per bunch crossing. For all signal and background samples we model the effect of the finite size of the beam spot by assuming the longitudinal 
 $z$-coordinate of the primary vertex follows a Gaussian distribution around the center of the detector, with a width of $45$\,mm~\cite{ATLAS:2010gaa}. This effect will be crucial for the high level trigger strategy presented 
 below. We neglect the width of the beamspot in the transverse direction.

Using a simplified simulation of the ATLAS inner tracker and calorimeter, we propagate all charged particles in the event in the magnetic field and compute the location of the hits, in order 
to estimate the detector response for both signal and background. This is necessary as energy loss and bremsstrahlung in the detector elements and/or service structures cannot be neglected 
for particles with $p_T \lesssim \text{few}\times 100$~MeV. At each intersection of a track with a detector element we compute the appropriate average 
energy deposition for the particle velocity and material in question. Moreover, whenever an electron traverses a detector element, a single bremsstrahlung photon may be radiated 
along the line of motion of the particle with a probability determined by the thickness of the element and with an energy deducted from the propagating charged particle. We finally compute the 
positions of these photons in the ATLAS ECAL, as well as the position of the all other particles that reach the calorimeter. 
A more detailed description of our detector simulation and assumptions is provided in Appendix~\ref{app:detecsim}.

\subsection{Analysis strategy}
\label{sec:AS}

\begin{figure}[t]
\includegraphics[width=0.45\textwidth]{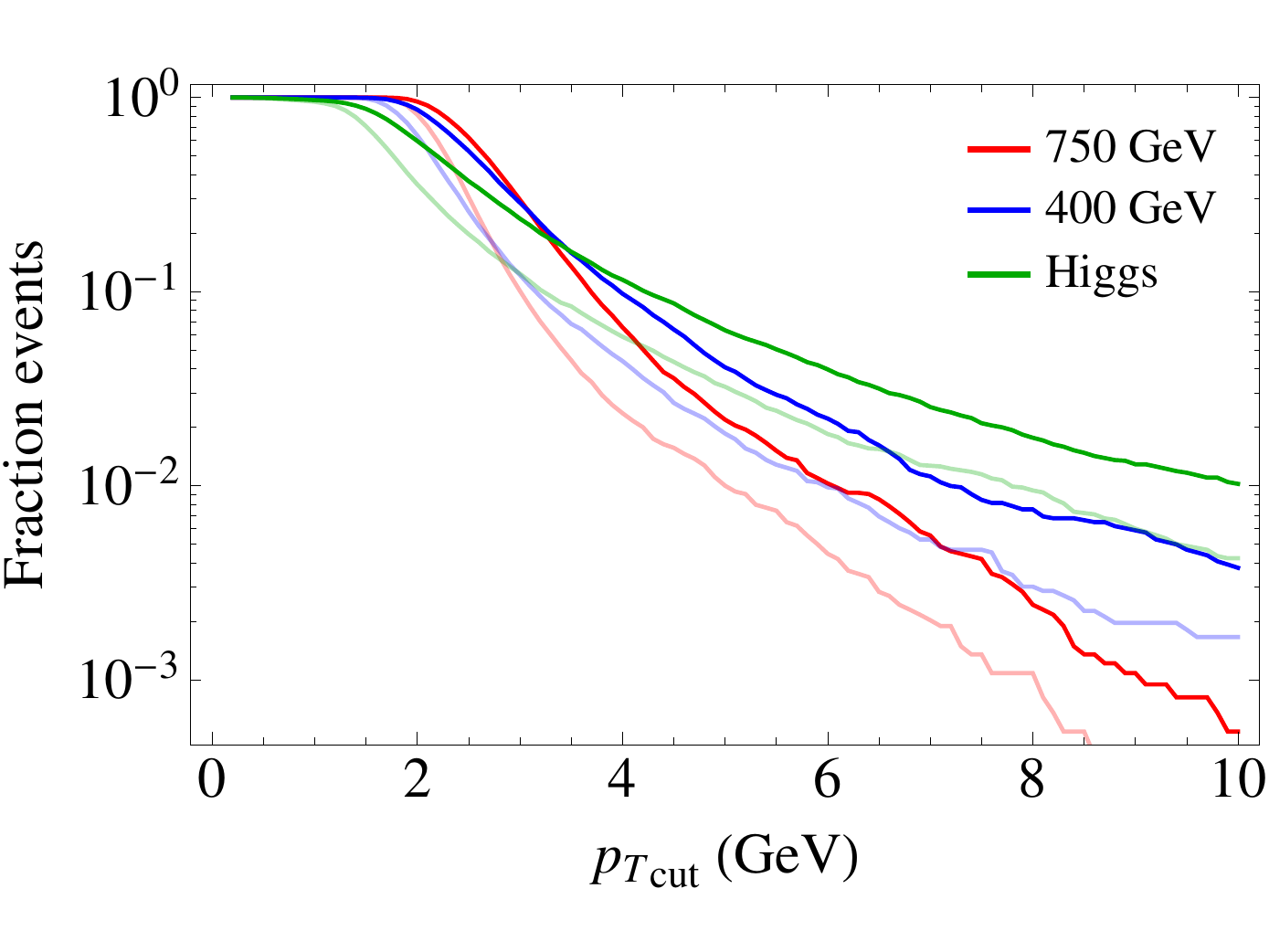}
\caption{Fraction of events with at least one (two) muons with a $p_T$ greater or equal the $p_{T_{\text{cut}}}$, for various benchmarks indicated by dark (light) curves.
Production through gluon fusion was assumed for the Higgs benchmark.}
\label{fig:hardparticle}
\end{figure}

Passing the L1 hardware trigger is the most challenging step for our signature, as no inner tracking information is available at L1. To be maximally conservative, we chose our benchmarks 
in Tab.~\ref{tab:sigbenchmark} such that hard muons are rare, which means that the standard muon triggers are not very efficient. 
This is quantified in Fig.~\ref{fig:hardparticle}, which shows the fraction of events with at least one muon passing a particular $p_T$-cut, as a function of the $p_T$ cut. 
We will therefore be forced to make use of other features in the events. 

\begin{figure}[t]
\includegraphics[width=0.5\textwidth]{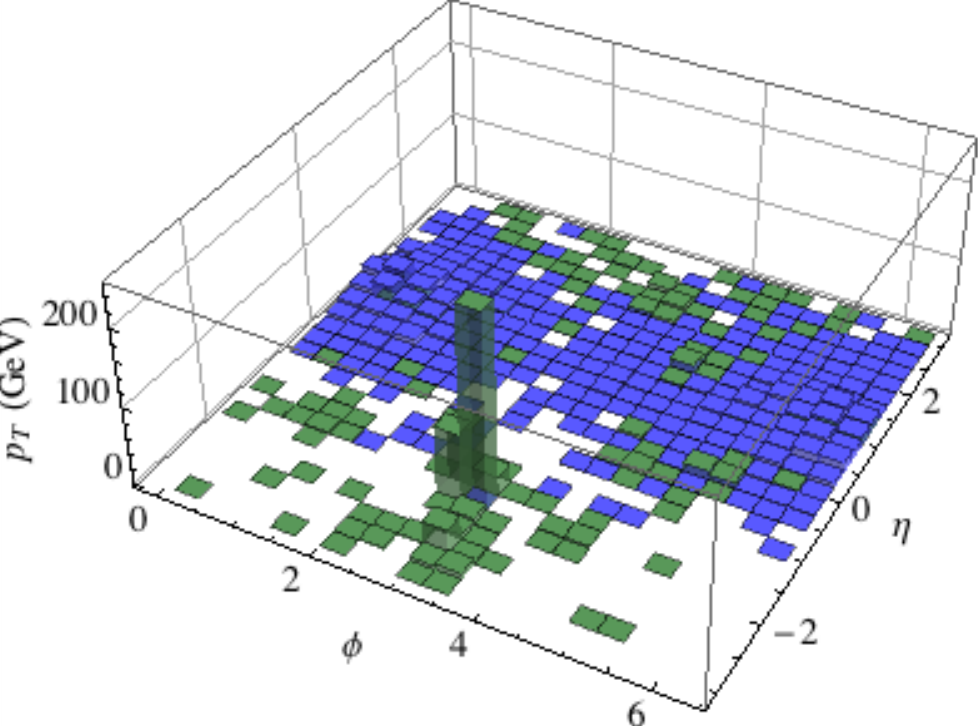}
\caption{Truth-level lego plot of an example event for the $m_\bomb=750$ GeV benchmark. The distribution of leptons (blue) and other particles (green) are 
indicated separately as function of the azimuthal angle and the pseudo-rapidity. This event contains a relatively hard ISR jet recoiling against the much more diffuse soft bomb. \label{fig:lego}}
\end{figure}

For very heavy soft bombs, $m_\bomb \gtrsim 1$~TeV, it may be possible to use a generic $H_T$ trigger on the scalar energy sum collected in the calorimeters. Since an $
\mathcal{O}(1)$ fraction of the energy is deposited in `loopers' that are too soft to reach the calorimeter,  such a trigger ceases to be efficient for lower masses.
For the benchmarks that we consider in Tab.~\ref{tab:sigbenchmark}, in the case of GF production one may instead trigger on event topologies with a sufficiently hard 
ISR jet. In the case of the VBF (VH) production channels for the Higgs benchmark, one may similarly trigger on a hard jet (jet or lepton), respectively. The presence of such hard objects further 
induces an asymmetry in the $p_T$ distribution of the soft bomb, as a relatively collimated object is recoiling against a 
large collection of soft particles. This is illustrated in Fig.~\ref{fig:lego} via a lego-plot of a sample truth-level event. Because a sizable fraction of the soft bomb particles are 
loopers that never reach the calorimeter, this imbalance will be also registered as missing transverse energy, $\MET$. This apparent missing energy 
signature, in combination with the jet or lepton itself, provides opportunities to pass L1 $\MET$, lepton or jet-based triggers, as analyzed in Sec.~\ref{sec:L1} below.

\begin{figure}[t]
\subfigure[~Inclusive \label{fig:ntrack}]{
\includegraphics[width=0.4\textwidth]{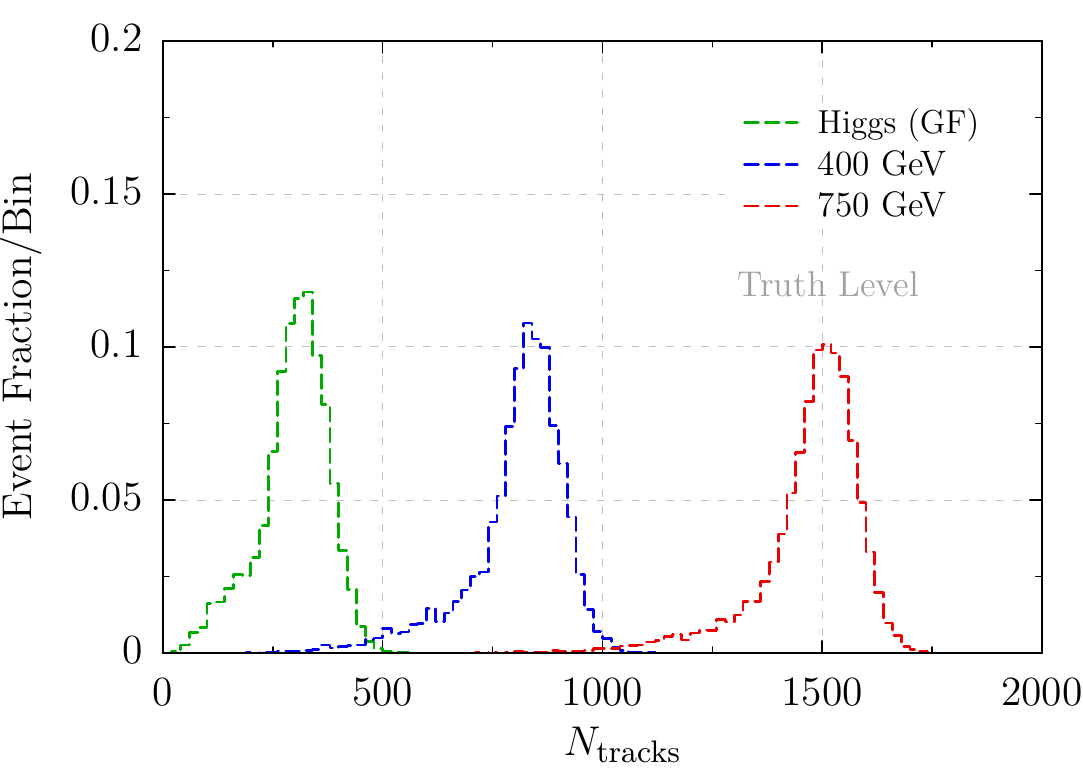}
}\hspace{1.5cm}
\subfigure[~$p_T>10$ MeV\label{fig:ntrack10}]{
\includegraphics[width=0.4\textwidth]{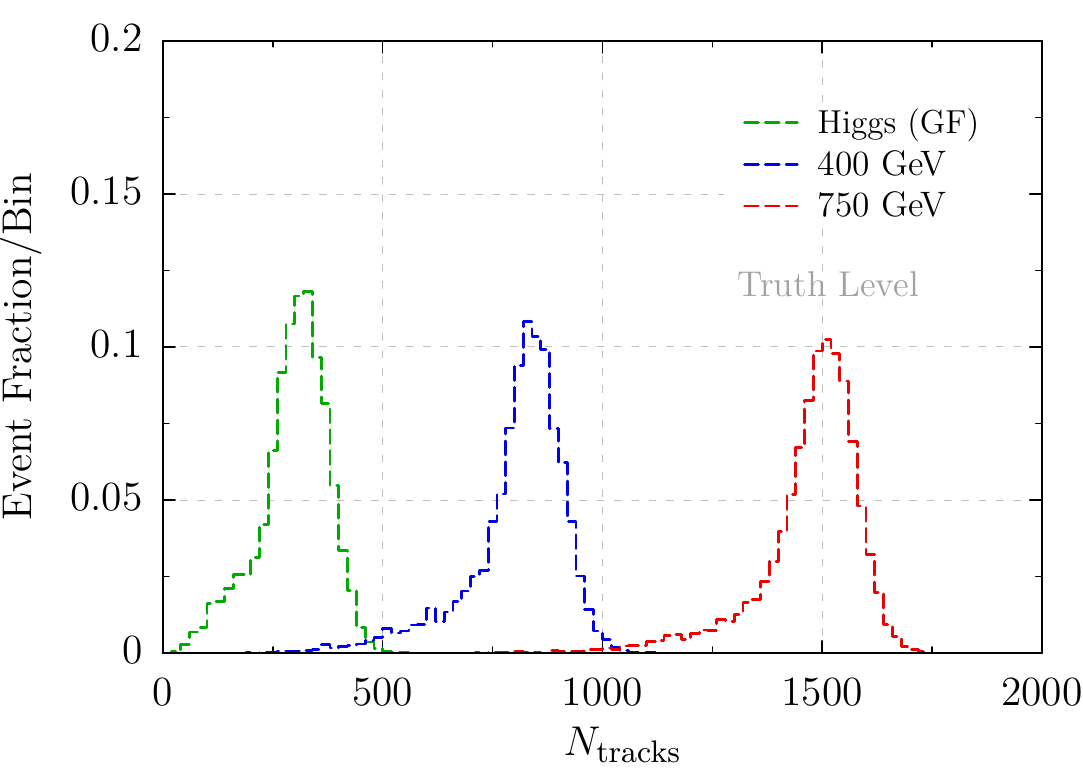}
}\\
\subfigure[~$p_T>400$ MeV\label{fig:ntrack400}]{
\includegraphics[width=0.4\textwidth]{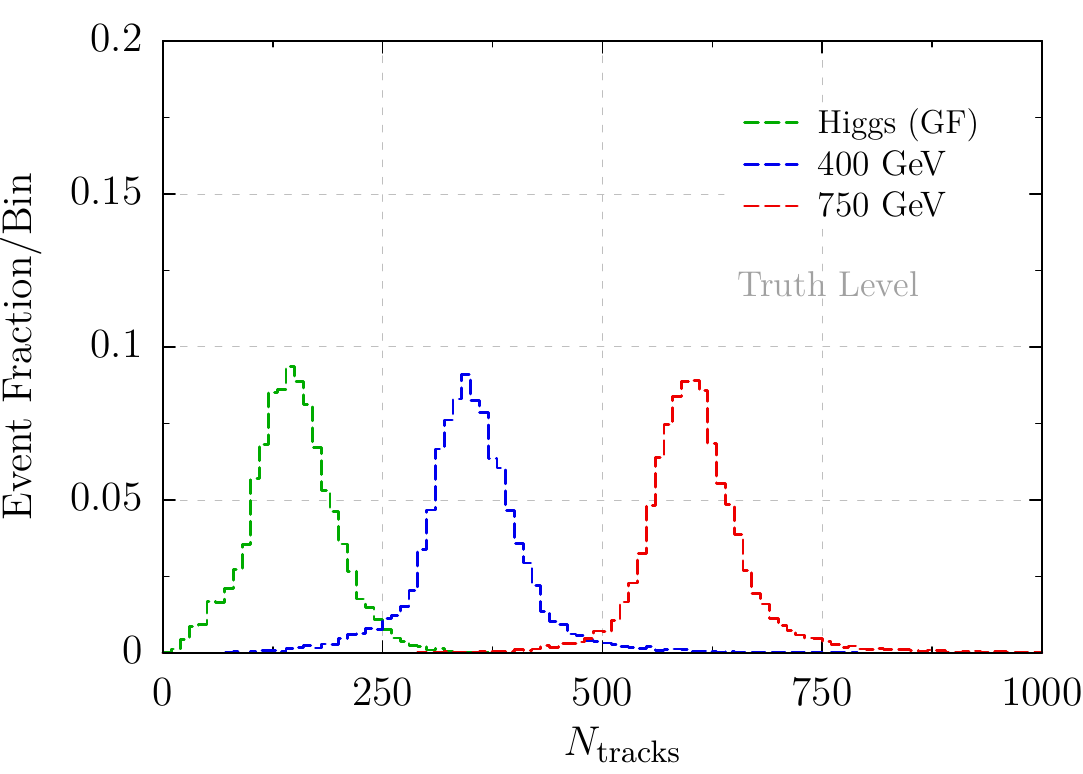}
}\hspace{1.5cm}
\subfigure[~$p_T>1$ GeV\label{fig:ntrack1000}]{
\includegraphics[width=0.4\textwidth]{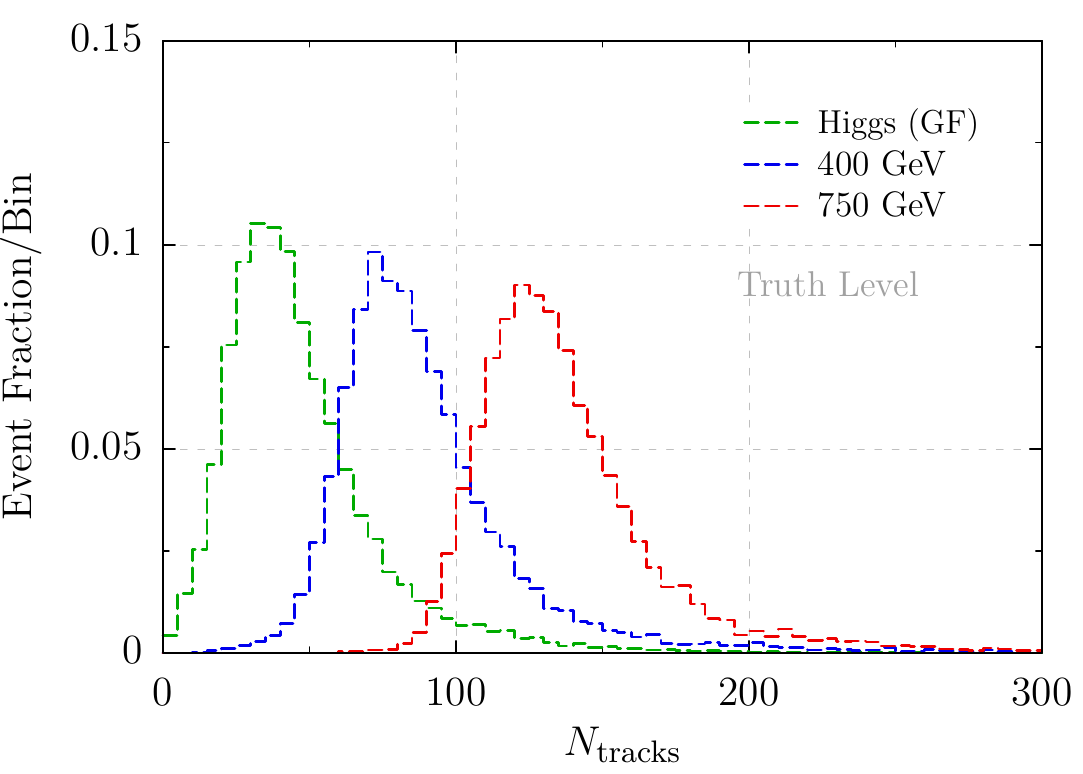}
}
\caption{Multiplicity of charged particles with $|\eta| < 2.4$ for the various signal benchmarks, with and without fiducial $p_T$ cuts.}
\label{fig:multiplicity}
\end{figure}

At the high level (software) trigger some forms of tracking information are available, however full track reconstruction is still not 
possible for every event. Instead we propose to directly use the distribution of the hits on the tracker surfaces, rather than tracks, to discriminate signal from background.\footnote{This approach somewhat resembles a multipole analysis that was proposed in the context of quirk annihilation~\cite{Harnik:2008ax}. However, the analysis in Ref.~\cite{Harnik:2008ax} 
was proposed as an off-line, even post-discovery, tool to discriminate quirks from a regular dijet resonance. Further, since the quirks mostly radiate to photons, this analysis made use 
of the ECAL rather than the inner tracker.}
In particular, we search for a signature in the form of a very concentrated, ring-shaped overdensity of hits, that is far more dense than the background produced by pile-up.
Such a `belt of fire' is most striking on the innermost layer of the tracker, as it will be sensitive even to the softest part of the signal spectrum. 
For ATLAS this layer is the Insertable B-Layer (IBL)~\cite{Capeans:1291633}, which can be reached by particles with $p_T \gtrsim 10$~MeV and $|\eta| < 2.4$. In Fig.~\ref{fig:multiplicity} 
we show the distribution of truth-level charged particle multiplicities for the three soft bomb GF benchmarks, with various fiducial $p_T$ cuts, and requiring $|\eta| < 2.4$. 
As a rule of thumb, the inclusive charged particle multiplicity in our benchmarks at truth-level is roughly $2 \times (m_{\bomb}/\text{GeV})$. The corresponding fiducial multiplicities are shown in 
Fig.~\ref{fig:ntrack10} for $p_T \gtrsim 10$~MeV, which corresponds to reaching the ATLAS IBL. As an alternative to our proposal, it may be possible to utilize the new ATLAS FTK system~\cite{Shochet:1552953} to get 
tracking information early in the trigger stream. This requires a $p_T$ greater than about 1 GeV, for which the corresponding multiplicities are shown in Fig.~\ref{fig:ntrack1000}.

\begin{figure}[t]
	\includegraphics[width=0.45\textwidth]{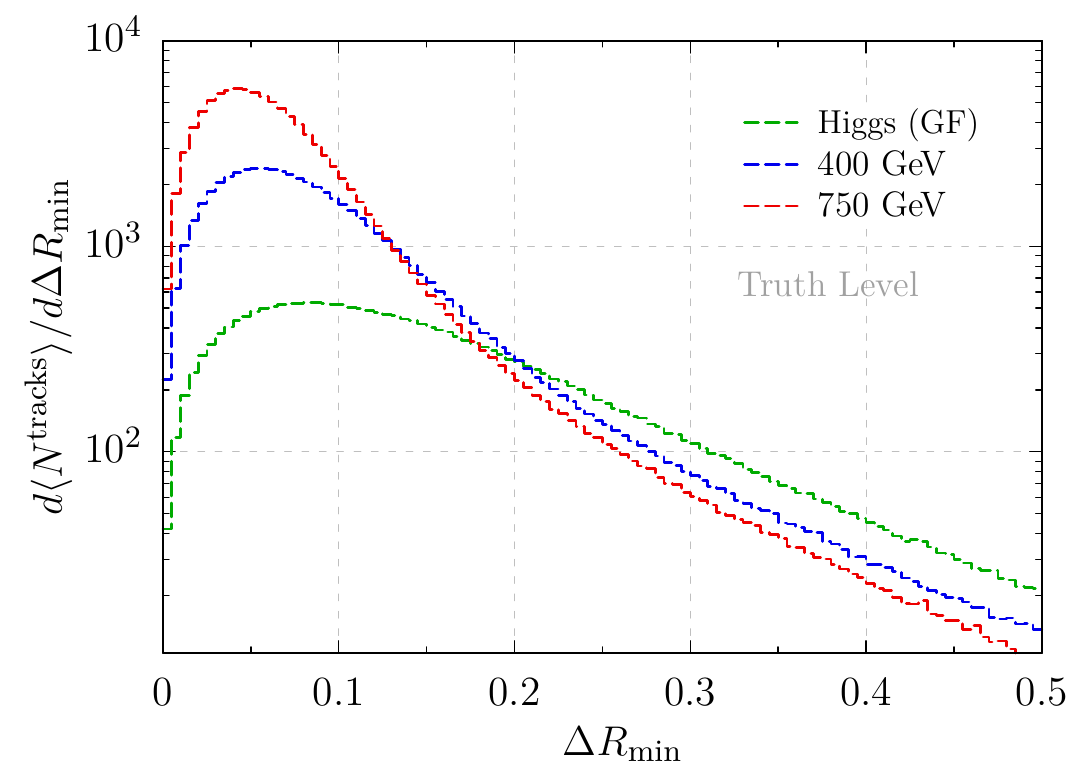}
	\caption{Distribution of minimum $\Delta R_{\rm min}$ over all soft bomb tracks.}
	\label{fig:MDR}
\end{figure}

Finally, once the event is written to tape, it will be possible to run the standard tracking algorithms to reconstruct those particles for which $p_T\gtrsim$ 400 MeV and $|\eta|<2.4$. 
Given the relatively large amount of tracks which pass these criteria, shown in Fig.~\ref{fig:ntrack400}, we expect the signal discrimination to be much further enhanced. 
However especially for the $m_\bomb=750$ GeV benchmark, one may be concerned that the density of tracks is too high for the reconstruction algorithm to function properly. 
Since the events are spherical in nature, this is not a major concern: In the tracking volume, one typically expects the distance in $\Delta R$ of each track to its nearest neighbor, 
$\langle \Delta R_{\text{min}} \rangle \sim \sqrt{2\pi 2|\eta|_\text{max}/ \langle N_{\text{tracks}} \rangle}/\pi$, where here $|\eta|  < 2.5$. Based on the multiplicities shown in 
Fig.~\ref{fig:multiplicity}, one then expects $\langle \Delta R_{\text{min}} \rangle \sim 0.13$, $0.08$, and $0.06$ for the Higgs, $400$ and $750$~GeV benchmarks, respectively.  
This is commensurate with the distributions shown in Fig.~\ref{fig:MDR}.  
For this degree for separation between the particles, one expects the track reconstruction to be nearly fully efficient~\cite{ATL-PHYS-PUB-2015-006}.

\subsection{Level 1 trigger}\label{sec:L1}
The ATLAS level one (L1) trigger at Run~II is implemented in hardware to reduce the total accepted rate to approximately $100$~kHz. 
For a soft bomb produced by gluon or vector boson fusion, the relevant L1 triggers are based on jet, multijet and $\MET$. For VH production, it is also possible 
to trigger on a muon from the associated vector boson. Given the sizable amount of energy in the soft bomb and the large population of electrons and positrons in each event, 
we also consider photon triggers, that search for clusters in the ECAL. At L1, electrons and photons are indistinguishable in the ECAL, so that hereafter the photon L1 trigger 
will also include contributions from electrons, and be referred to as an EM trigger.

\begin{figure}[t]
	\includegraphics[width = 0.45\linewidth]{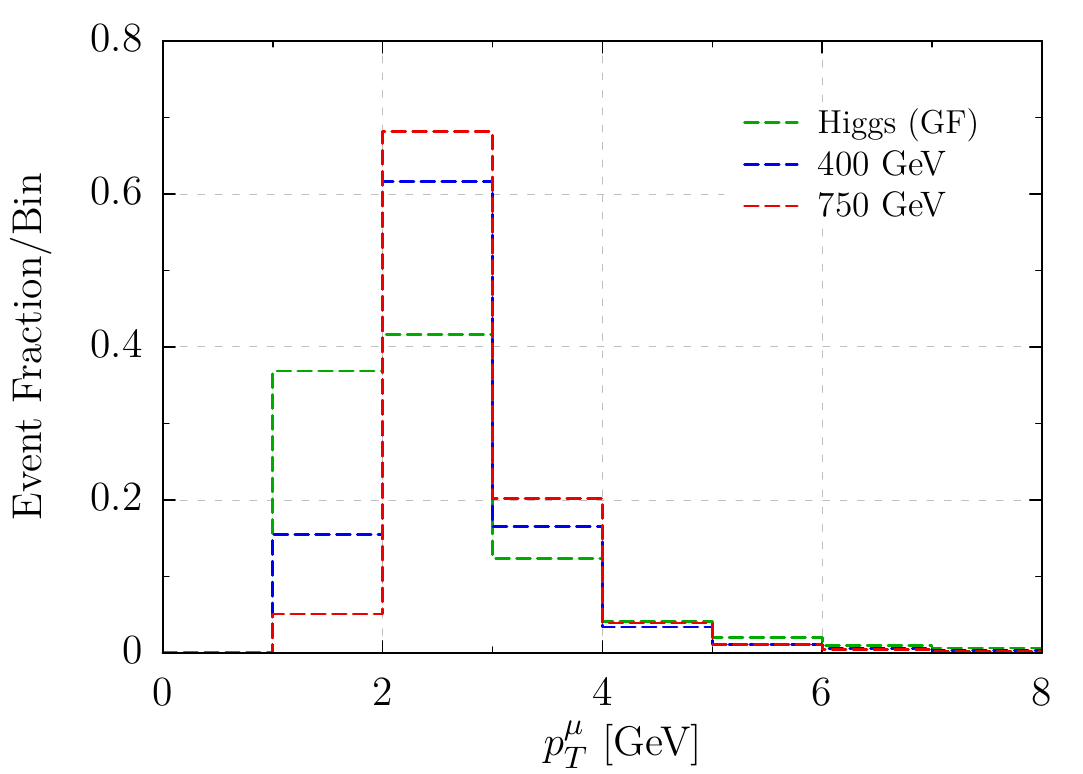}\hfill
	\includegraphics[width = 0.45\linewidth]{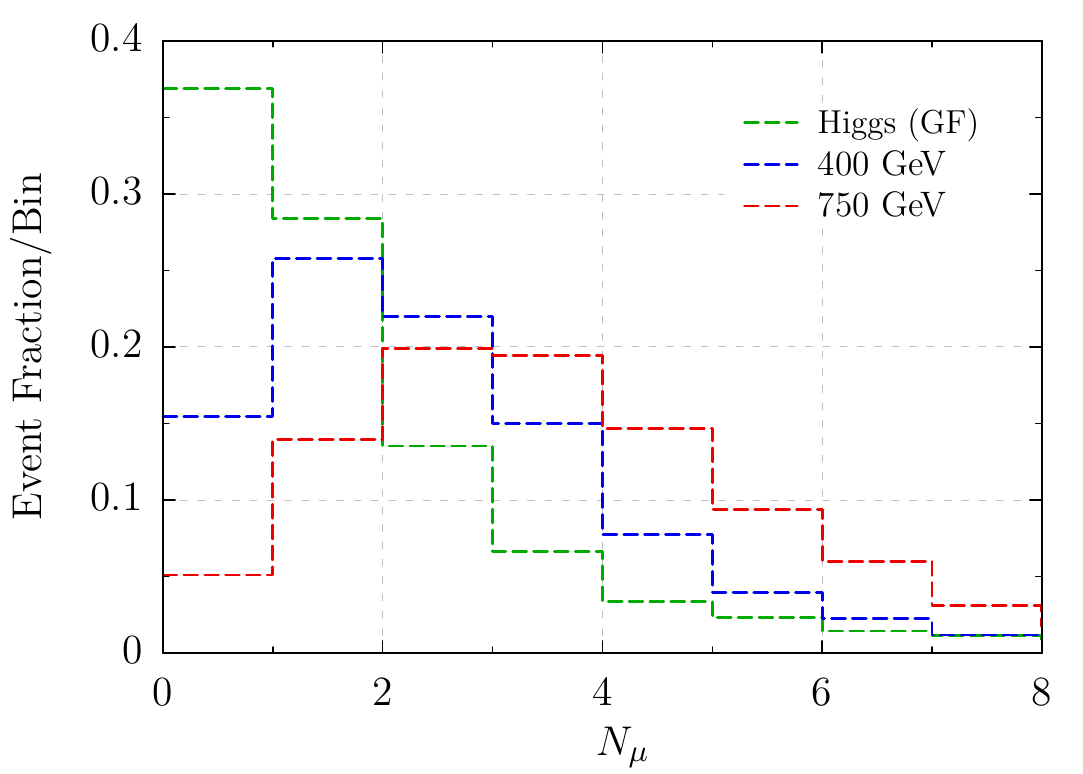}
	\caption{Left: $p_T$ distribution of the hardest muon, after propagation through the inner detector, with $|\eta| < 4.9$. Right: Number of muons with $p_T > 2$~GeV and $|\eta| < 4.9$, 
	that escape the calorimeters.}
	\label{fig:pTmu}
\end{figure}

Our analysis of soft bomb events after propagation, including computation of trigger efficiencies, is performed with \texttt{ATOM 0.9}~\cite{Atom}. 
Since the trigger efficiencies presented in~\cite{ATL-DAQ-PUB-2016-001} are expressed in terms of offline reconstructed objects, we use corresponding offline properties for the detector response in defining jets, leptons, photons and $\MET$.
Jets are clustered with FastJet~\cite{Cacciari:2011ma} using the anti-$k_t$ jet algorithm and a jet distance parameter $R = 0.4$, as an approximation and in lieu of ATLAS's 
sliding-window jet algorithm used at L1~\cite{ATLAS:2016qun}. (The SIS Cone algorithm~\cite{Salam:2007xv} is a possible alternate choice, expected to yield 
similar results.) Clustering is performed over the full $-4.9 < \eta < 4.9 $ range of the ATLAS detector. Given the large multiplicity of particles in the soft bomb
and its diffuse nature, pile-up mitigation algorithms may bias the hardness of the reconstructed jets and $\MET$. Therefore we implement event-by-event pile-up subtraction using
jet areas and a energy density estimated from $k_t$ jets of the same radius, which is the ATLAS offline pile-up removal prescription.
For the L1 analysis we include jet energy resolution smearing as in Ref.~\cite{ATLAS-CONF-2015-037}. 
The calorimeter $\MET$ is computed from the transverse momentum of all visible particles that reach the calorimeters. This 
includes the vast majority of the soft bomb muons, which typically have insufficient $p_T$ to escape the hadronic calorimeter and be detected in the muon chambers. We take 
the muon $p_T$ threshold for inclusion in calorimeter MET to be $p_T < 2$~GeV. From Fig.~\ref{fig:pTmu}, we see that a majority of 
events do have several muons exceeding this threshold, which correspondingly alters the $\MET$ in such bombs by several GeV.  

Photons and electrons are required to have a minimum $E_T > 5$~GeV in order to be retained for the L1 analysis. To account for the merging of adjacent photons and electrons into a single calorimeter cluster,
we clustered photons with radius $R = 0.05$ via the Cambridge/Aachen algorithms. We also cluster photons around electrons/positrons using a fixed cone centered on the lepton with the same $R = 0.05$ radius, in lieu of the specific 
clustering algorithms used at L1. In treating the electron and photon contributions together, we perform overlap removal of photons near electrons by requiring a minimum separation $R=0.05$, in order to avoid double-counting.
Detector energy resolution effects for electrons and photons are included~\cite{Aad:2014nim,ATLAS:2011kuc}, though these effects are expected to be subleading compared to errors arising from the approximate treatment of bremsstrahlung effects 
in our inner detector simulation (see App.~\ref{app:detecsim}). We also apply isolation criteria to photons and electrons, requiring an isolation cone of outer (inner) radius $R = 0.2$ 
($R = 0.05$) and an isolation threshold of $2$~GeV.  The isolation threshold requirements can be difficult to satisfy, given the typically high multiplicity of soft electrons, positrons and 
photons reaching the ECAL in a soft bomb event. However, for the Higgs benchmark, with correspondingly lower particle multiplicities, photons and electrons can be become more important for 
the VH production channel. Finally, muons are retained for the L1 analysis provided they have $|\eta| < 2.4$ and a minimum 
$p_T > 2$~GeV, such that they reach the muon chambers. This is mostly relevant for VH production.

\begin{figure}[t]
	\includegraphics[width = 0.49\linewidth]{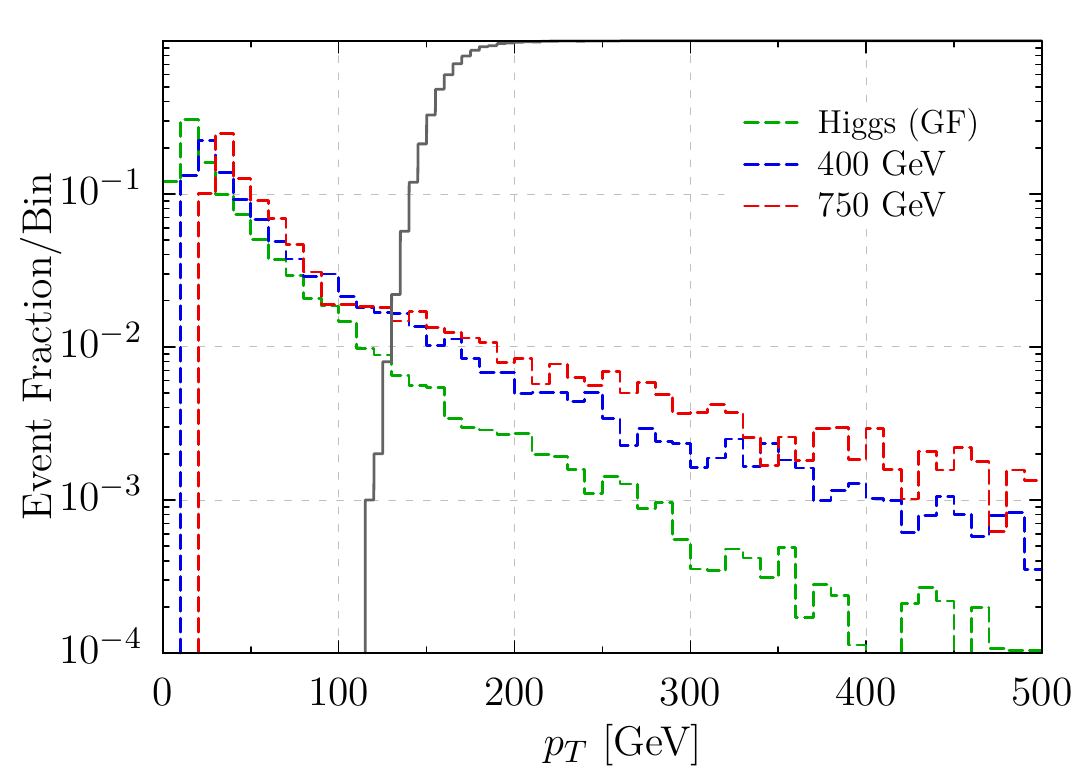}\hfill
	\includegraphics[width = 0.49\linewidth]{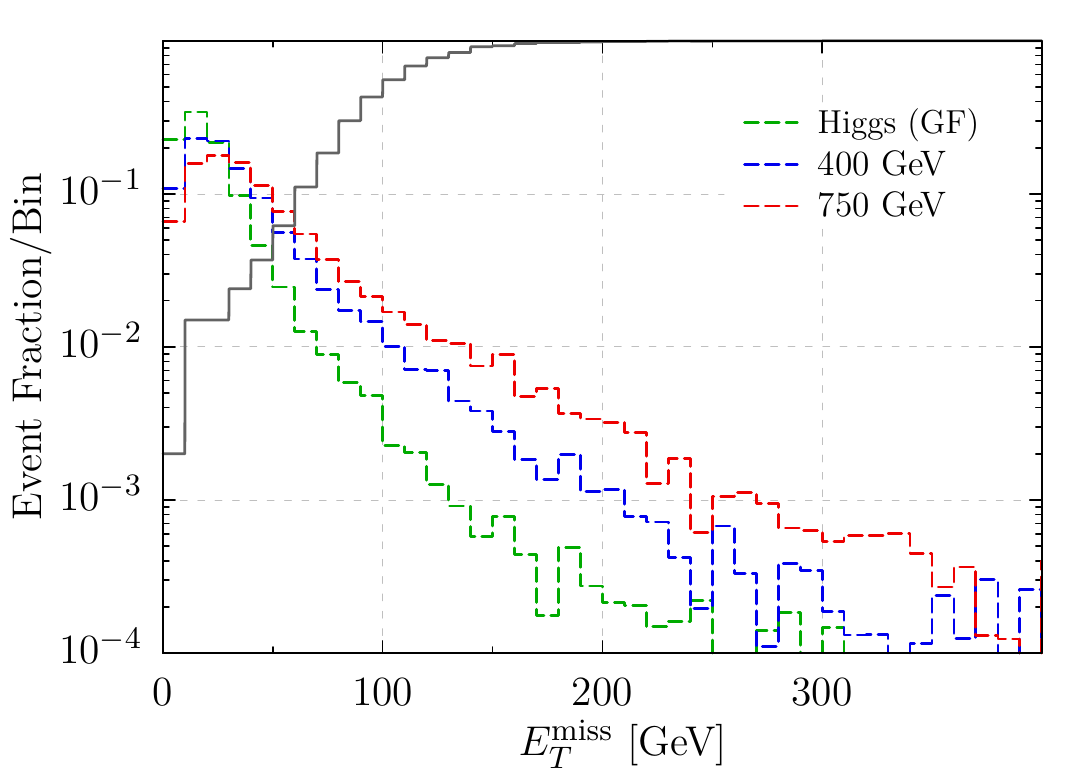}
	\caption{Distributions for $p_T$ of hardest jet (left) and $\MET$ (right), for the soft bomb GF
benchmarks. Also shown are applicable turn-on efficiency curves (gray)~\cite{ATL-DAQ-PUB-2016-001}, with the same vertical scale as the corresponding distributions. 
These turn on curves are applied over the full $|\eta| < 4.9$ range.}
	\label{fig:pTMET}
\end{figure}

The relevant 2015  L1 trigger thresholds~\cite{ATL-DAQ-PUB-2016-001,Aaboud:2016leb} are reproduced in the left column of Table~\ref{tab:L1E}. Our objects have been reconstructed at truth-level 
and corrected with offline energy resolution smearing effects. Given that both the energy scales and energy resolutions differ between L1 and offline objects, we account for these 
discrepancies by using the corresponding publicly available turn-on efficiency curves\footnote{Wherever a curve is not available for a given cut threshold, we rescale the closest available one. In multi-object triggers, we 
also assume that the turn-on curves for each object are uncorrelated, a reasonable assumption for well-separated objects.}. In Fig.~\ref{fig:pTMET} we display the hardest jet $p_T$ and $\MET$ distributions for GF 
production, along with the corresponding turn-on efficiencies for both triggers.  The tail of $\MET$ distribution drops exponentially, while the tails 
of the jet $p_T$ distributions are longer, and the turn-on efficiency curve somewhat steeper, so that this trigger is always more efficient than $\MET$ for 
the GF benchmarks. 

In the remainder of Table~\ref{tab:L1E} we show the trigger efficiencies extracted from a sample of $10^4$ soft bomb events. We also 
compute a `combined' trigger efficiency for a bomb to pass at least one of these L1 triggers, expecting that the individual triggers may not be 
fully correlated over all the events. Turn-on effects are included for the single jet, multijet, $\MET$, lepton and EM triggers. 
For the combined trigger efficiency, we neglect possible correlations between the different underlying turn-on curves. For the Higgs portal benchmarks we also show the production cross section times the trigger efficiencies. Although the efficiency for the GF mode is comparatively low, it is still the most sensitive channel due to the higher inclusive cross section. 
The VH and VBF channels nevertheless produce non-negligible contributions to the accepted L1 rate.

\begin{table}[t]
\renewcommand*{\arraystretch}{1.1}
\newcolumntype{C}{ >{\centering\arraybackslash} m{2.85cm} <{}}
\scalebox{0.8}{
\parbox{1.2\linewidth}{
\begin{tabular}{|c|C|C|C|C|C|}
	\hline
							& Higgs (VH) 	& Higgs (VBF)	& Higgs (GF) & $400$~GeV (GF) & $750$~GeV (GF)\\
	\hline\hline
	1j ($p_T > 100$~GeV)		& $0.057 \pm 0.002$ 	& $0.107 \pm0.003$		&$0.034 \pm 0.001$ 		& $0.104 \pm 0.002$  	& $0.169 \pm 0.002$ \\
	3j ($p_T > 40$~GeV)		& $0.026 \pm 0.001$ 	& $0.045 \pm 0.002$		& $< 1\%$ 			& $< 1\%$ 			& $0.019 \pm 0.001$ \\
	4j ($p_T > 20$~GeV)		& $0.019 \pm 0.001$		& $0.034 \pm 0.002$		& $< 1\%$ 			& $0.016 \pm 0.001$  	& $0.052 \pm 0.002$ \\
	$\MET > 50$~GeV 			& $0.088 \pm 0.002$		& $0.063 \pm 0.001$		& $0.030\pm 0.001$ 		& $0.077 \pm 0.001$ 	& $0.136 \pm 0.002$ \\
	1$\gamma/e$ ($E_T > 20$~GeV) & $0.045 \pm 0.002$	& $0.011 \pm 0.001$		& $< 1\%$ 			& $< 1\%$ 			& $< 1\%$  \\
	1$\mu$ ($p_T > 20$~GeV) 	& $0.073 \pm 0.002$		& $0.011\pm 0.001$		& $< 1\%$ 			& $< 1\%$  			& $< 1\%$ \\
	Combined					& $0.224 \pm 0.003$		& $0.162 \pm 0.003$		& $0.055 \pm 0.001$ 	& $0.140 \pm 0.002$ 	& $0.217 \pm 0.002$\\
	\hline
	$\sigma_{pp\rightarrow h +X}\times \epsilon_{\mathrm{comb}}$ (pb)	& $0.50$ & $0.60$ & $2.39$ & -- & -- \\
	\hline	
\end{tabular}
}}
\caption{L1 efficiencies for the different benchmarks and trigger paths, where the quoted uncertainties are statistical only. 
For the Higgs portal benchmark, the last row indicates production Higgs cross section~\cite{Heinemeyer:2013tqa} multiplied with combined L1 trigger efficiency ($ \epsilon_{\mathrm{comb}}$). 
}
\label{tab:L1E}
\end{table}

In the specific benchmark chosen here, many of the jets originating from the bomb will have an anomalously high EM energy fraction, while at the same time, they may be too wide to be considered for isolated photon or electron reconstruction. As such they may fail noise-cleaning requirements. This is not, however, necessarily an issue because if they are discarded they will instead increase the likelihood of passing the $\MET$ trigger. It may also be in principle possible to modify such requirements to retain these jets, and, moreover, this issue is very specific to the $\etav$ leptonic decay mode chosen here, and will disappear as soon as decays into pions are included. Therefore we believe that the results presented in Table~\ref{tab:L1E} provide a realistic estimate of achievable L1 efficiencies.

\subsection{High level trigger}\label{sec:HLT}
The ATLAS HLT presently outputs a total rate of 1 kHz, which is roughly a 30\% 
improvement with respect to Run~I~\cite{ATL-DAQ-PUB-2016-001,Aaboud:2016leb}. While the L1 trigger only has access to the calorimeter and the muon 
chambers, the HLT can make use of the inner detector as well. In particular tracking is possible at the HLT, but with some important caveats:
Since tracking is a fairly time consuming step and does not scale linearly with the instantaneous luminosity, it is only possible to run the 
algorithm on a small subset of events passing the L1 trigger. Alternatively, one may choose to process more events by only considering a 
small region of interest in the detector, which is seeded by the L1 trigger. ATLAS tracking capabilities on the trigger level will be further 
enhanced when the new FTK system comes online~\cite{Shochet:1552953}, although the rather soft $p_T$ spectrum expected in these events
may still pose a severe challenge. As mentioned in Sec.~\ref{sec:strategy}, it should be feasible to fully reconstruct soft bomb events off-line. However, due to the 
daunting combinatorics in the reconstruction of such extremely busy events, this approach is likely to be too time consuming to 
implement at the HLT. Even if it were feasible, we now proceed to show that it is possible to bypass track reconstruction entirely, 
while still maintaining a good signal efficiency. 

The key idea for our HLT analysis is to design a discriminating variable which can be applied directly on the 
 \emph{hits} in tracker. There are, however, a number of experimental 
subtleties with this approach as well. Firstly, $\delta$-rays and particles with small angle of incidence
tend to light up multiple neighboring pixels. To address this issue, ATLAS constructs clusters of hits if at least two pixels in the same 3$
\times$3 grid light up, and recursively keeps adding hits to the cluster until all neighboring hits are accounted for.  For the heavy $m_{\bomb}
=750$ GeV benchmark, neighboring hits occur on average for  only 0.4\% of all hits in the soft bomb, and for the purpose of our analysis 
we therefore identify the hits from our simulation with the reconstructed clusters in the experiment. We elaborate further on this subtlety in Appendix 
\ref{app:detecsim}. Secondly, as particles spread out in the detector, the hits from the bomb 
tend to be more and more diffuse for layers further away from the collision point. For this reason the innermost layer of the ATLAS tracker -- the IBL -- is the most 
sensitive to the soft bomb signature, using our approach. The IBL starts at just 31 mm from the interaction point with a pixel-size of $50\times 250\;{\rm \mu m}^2$. It extends 32 cm is each 
direction along the $z$-axis, which corresponds to $|\eta|<2.58$.

\begin{figure}[t]
\includegraphics[width=0.45\textwidth]{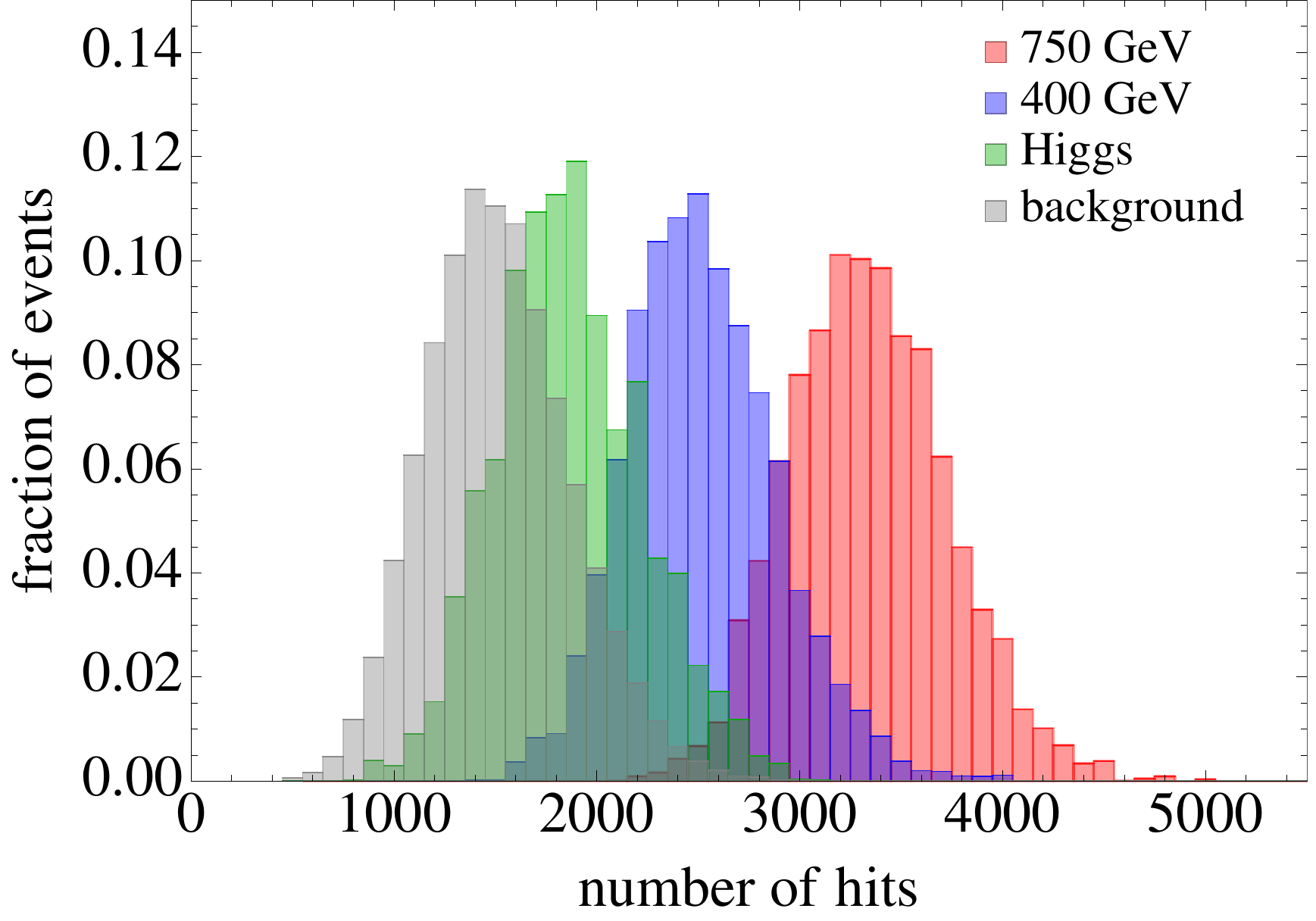}
\hfill
\includegraphics[width=0.45\textwidth]{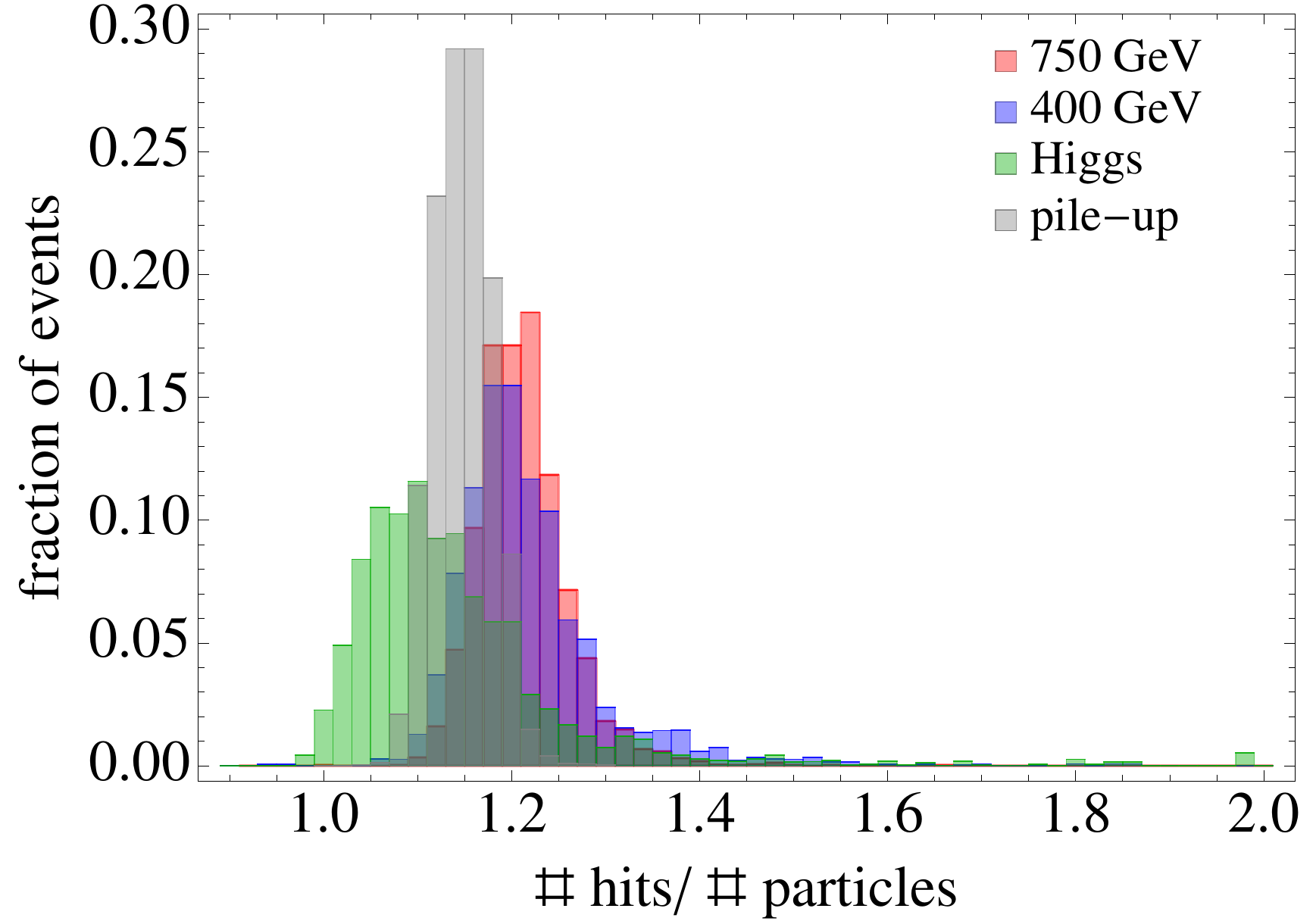}
\caption{Left: Number of hits on the ATLAS IBL for signal and background, both including a pile-up contribution corresponding to $\langle
\mu\rangle=$50. Right: Ratio of the number of hits on the ATLAS IBL over the number of tracks with $|\eta|<2.5$. The signal samples do 
not include pile-up in this plot. \label{fig:nrhits}}
\end{figure}

Figure~\ref{fig:nrhits} shows the distribution of the number of hits expected on the IBL, as generated by our MC simulation. The left hand panel indicates the absolute 
number of hits for the signal benchmarks in Tab.~\ref{tab:sigbenchmark}, compared to the background from dijets. Both for signal and dijet background, contributions from pile-up events are 
included, assuming an average number of pile-up vertices $\langle\mu\rangle=50$, which is roughly a factor of 2 higher than the 2016 running conditions. Observe that the distribution of hits 
is non-Poissonian for all samples, because the number of charged particles from minimum bias events within the 
acceptance of the tracker is itself strongly non-Poissonian. The right hand panel of Fig.~\ref{fig:nrhits} shows the average number of hits per particle within the tracker volume, separately 
for pile-up and for the signal. For all samples the average number of hits per tracks is close to one, which indicates that the 
majority of tracks either reach the calorimeter or become loopers only in the outer part of the tracker. 

The main challenge for our proposed triggering strategy is to separate the soft bomb signal from pile-up and $Z+$jets/di-jets backgrounds as efficiently as possible. Assuming that before cuts the 
signal rate is much smaller than the   background rate, $S/B \ll1$, we emphasize that it is sufficient here to reject the background contributions only to the extent that the background rate 
becomes small compared to the total HLT 
output rate. Depending on how much bandwidth one would want to dedicate to this trigger, this implies a desired background rejection 
level of the order of $10^{-3}$ to $10^{-4}$. The $S/B$ ratio may still remain quite small after this background rejection. However, once the events are written to tape, $S/B$ can be further 
improved by fully reconstructing the events off-line. 

From the left hand panel of Fig.~\ref{fig:nrhits} we see that the $750$~GeV benchmark, and to a lesser extent the $400$ GeV benchmark, can 
already be separated from the background, simply by counting the total number of hits. The Higgs portal benchmark on the other hand 
typically only produces a few hundred hits, which is a relatively small perturbation on top of the pile-up contribution. This makes it difficult 
to separate from the background by counting only. 

\begin{figure}[t]
\subfigure[~event display]{
\includegraphics[width=0.31\textwidth]{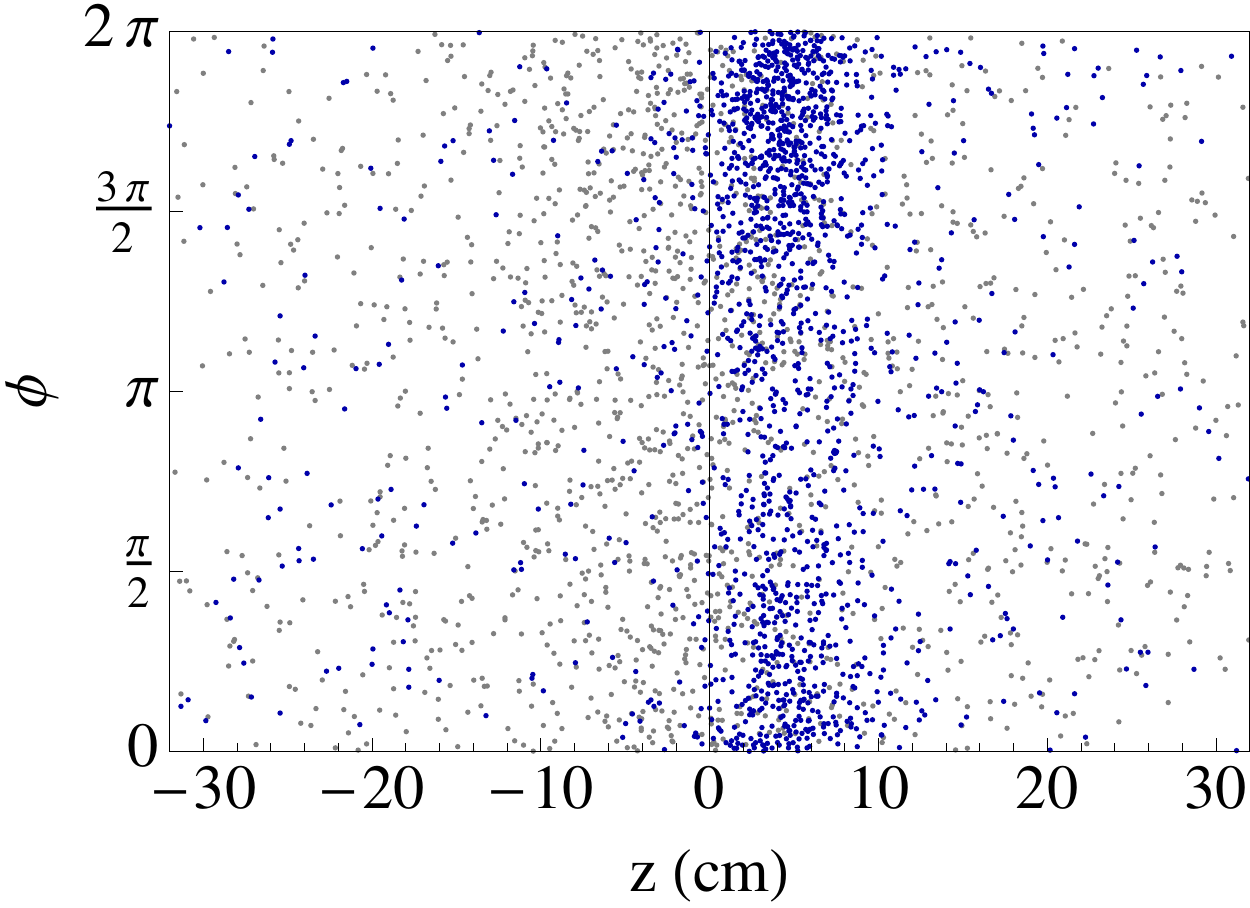}
}
\subfigure[~$z$ distribution\label{fig:trackereventz}]{
\includegraphics[width=0.31\textwidth]{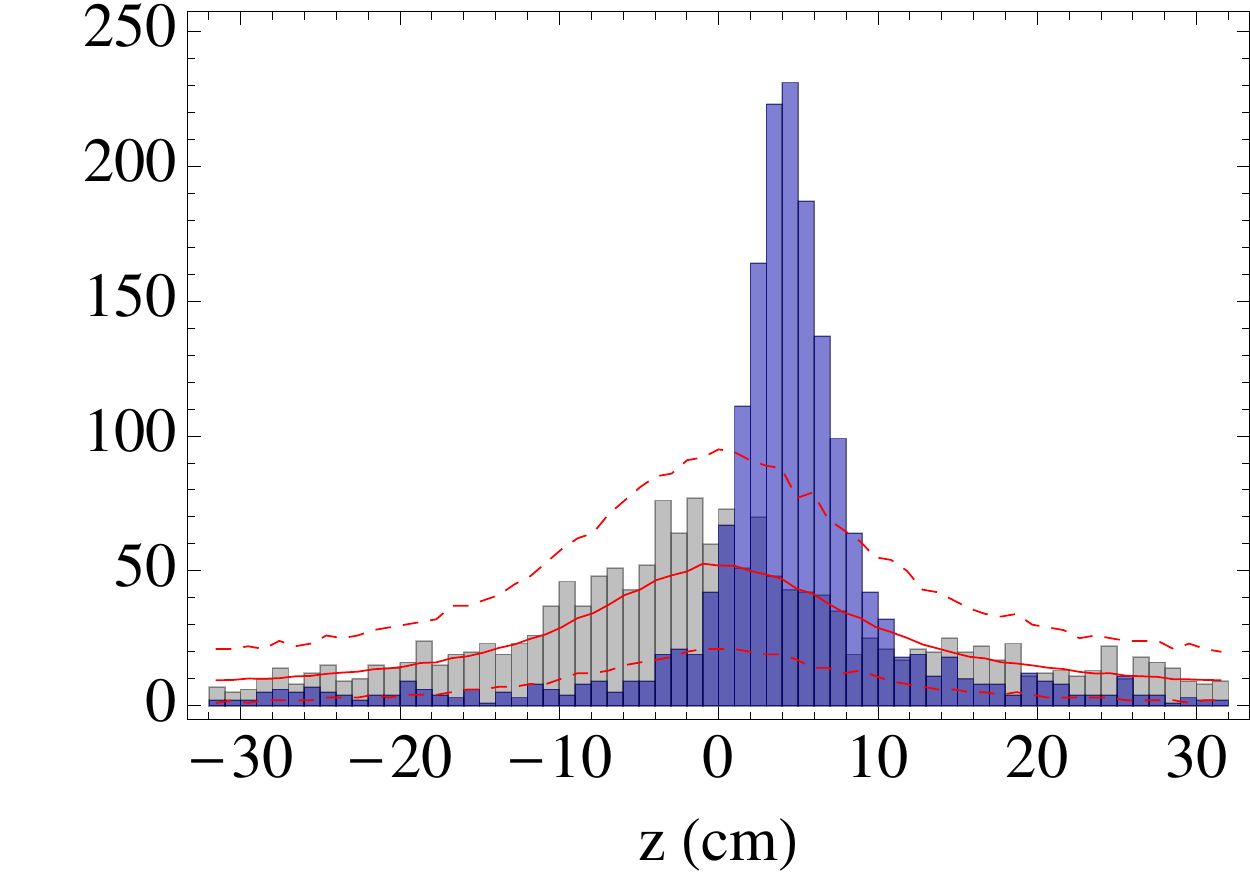}
}
\subfigure[~$\phi$ distribution\label{fig:trackereventphi}]{
\includegraphics[width=0.31\textwidth]{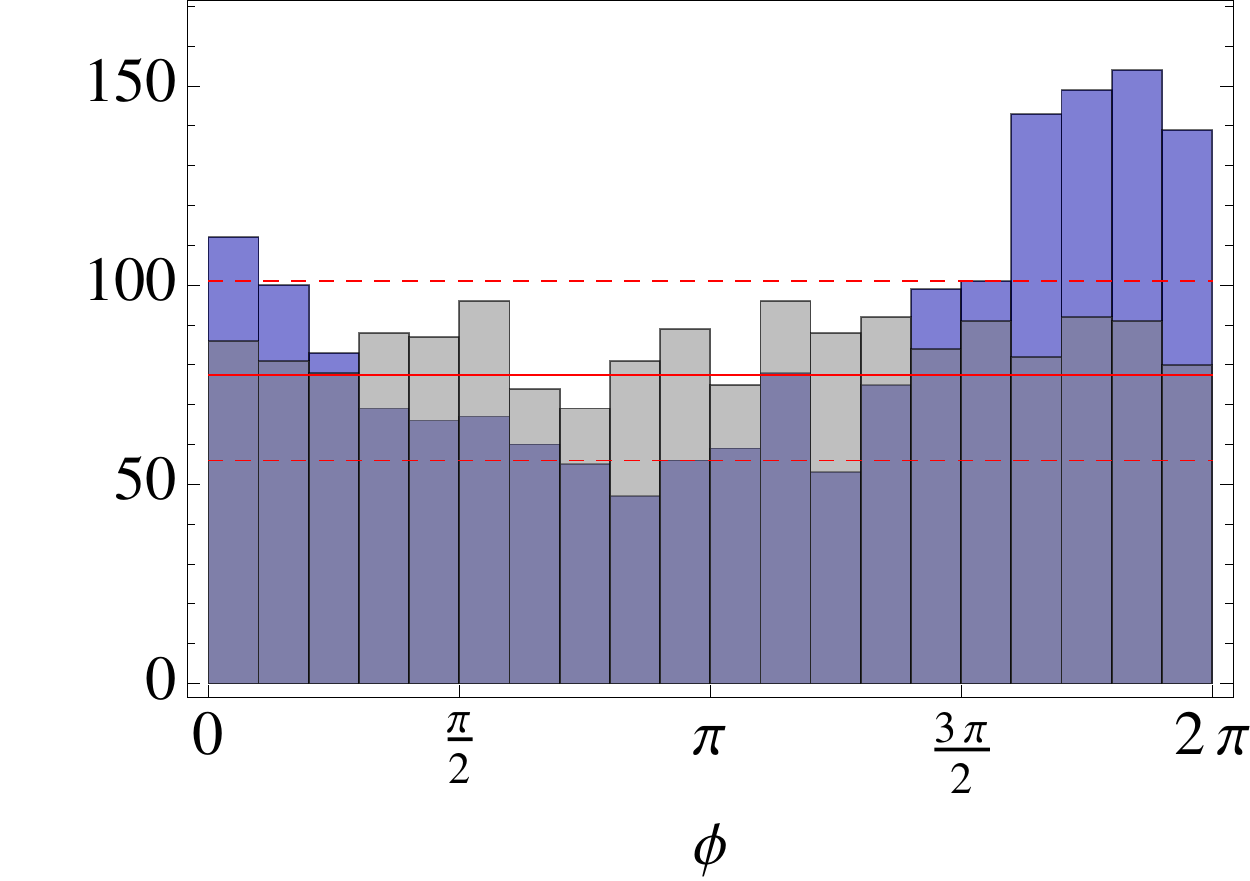}
}
\caption{\label{fig:trackerevent}Distribution of the hits on the ATLAS IBL for an example signal event of the $m_{\bomb}=750$ GeV 
benchmark. Hits from the primary vertex (pile-up) are marked in blue (gray). Solid (dashed) red lines indicate the bin-by-bin expectation 
value (99\% interval) for pile-up only events.}
\end{figure}

However it is possible to further improve the discriminating power by also accounting 
for the spatial distribution of the hits. In Fig.~\ref{fig:trackerevent} we show the longitudinal $z$ and axial $\phi$ distributions of tracker hits for a sample soft bomb event, compared to 
the pile-up contributions. To roughly indicate the size of the fluctuations in the pile-up sample, we include the bin-by-bin  99\% CL envelope for the pile-up distribution.
The soft bomb signal tends to produce a highly localized `\emph{belt of fire}' near the primary interaction point, while pile-up hits arise from a larger number of longitudinally separated vertices, 
and are therefore more uniformly spread out in the tracker. The pile-up hits also tend to be axisymmetric around the 
beam axis of the detector, and parity symmetric around the detector center. The longitudinal location of the soft bomb interaction point can be slightly off-center, due to non-zero longitudinal 
size the beam spot.
The bulk of the hits from a soft bomb are therefore generally somewhat displaced from the center in the z-direction. (We emphasize that we took all decays to be prompt, and this effect is 
therefore unrelated to the longitudinal and transverse displacements one expects for displaced decays.) The soft bomb events moreover tend to be somewhat asymmetric in $\phi$, as the L1 
trigger demands that the bomb is recoiling against an ISR jet or associated vector boson.

In order to obtain a quantitative estimate for the 
discriminating power -- the background rejection power -- of this method, we now present two different discrimination variables.  Both only 
include the IBL hits and marginalize over $\phi$: Including the non-trivial $\phi$ dependence and the remaining layers of the tracker will
further enhance the sensitivity, although probably at the expense of additional computation time. In our estimates of the efficiencies, we account for correlations between the L1 and HLT triggers, which we find not to be very significant for the variables studied.
The hard scattering event in the background sample
only produces a small number of hits compared to pile-up, and the background events can therefore essentially be viewed as pile-up only events in this part of the analysis.

\subsubsection{Fisher discriminant}
\begin{figure}[p]\centering
\includegraphics[width=0.4\textwidth]{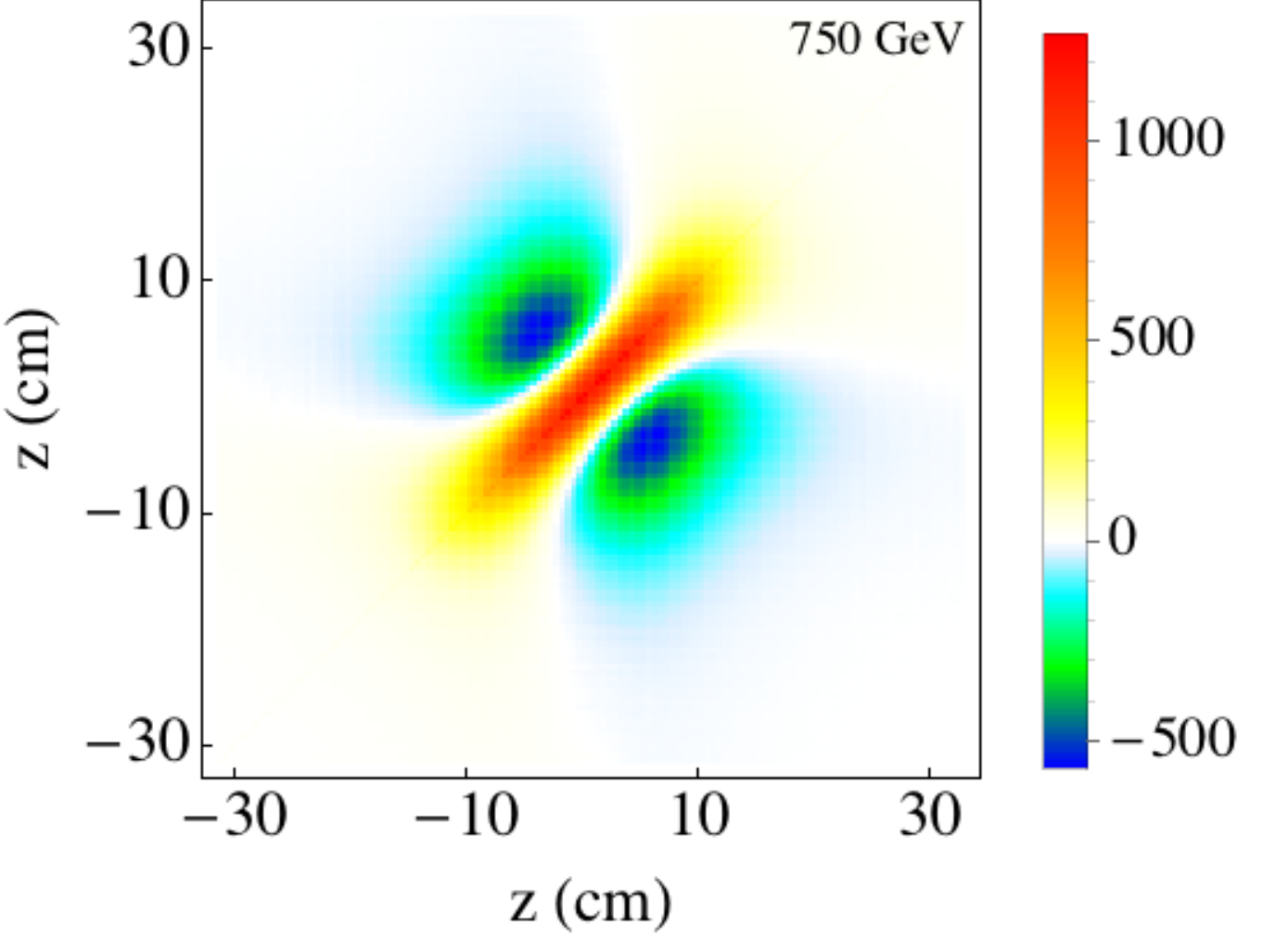}\hspace{1.5cm}
\includegraphics[width=0.4\textwidth]{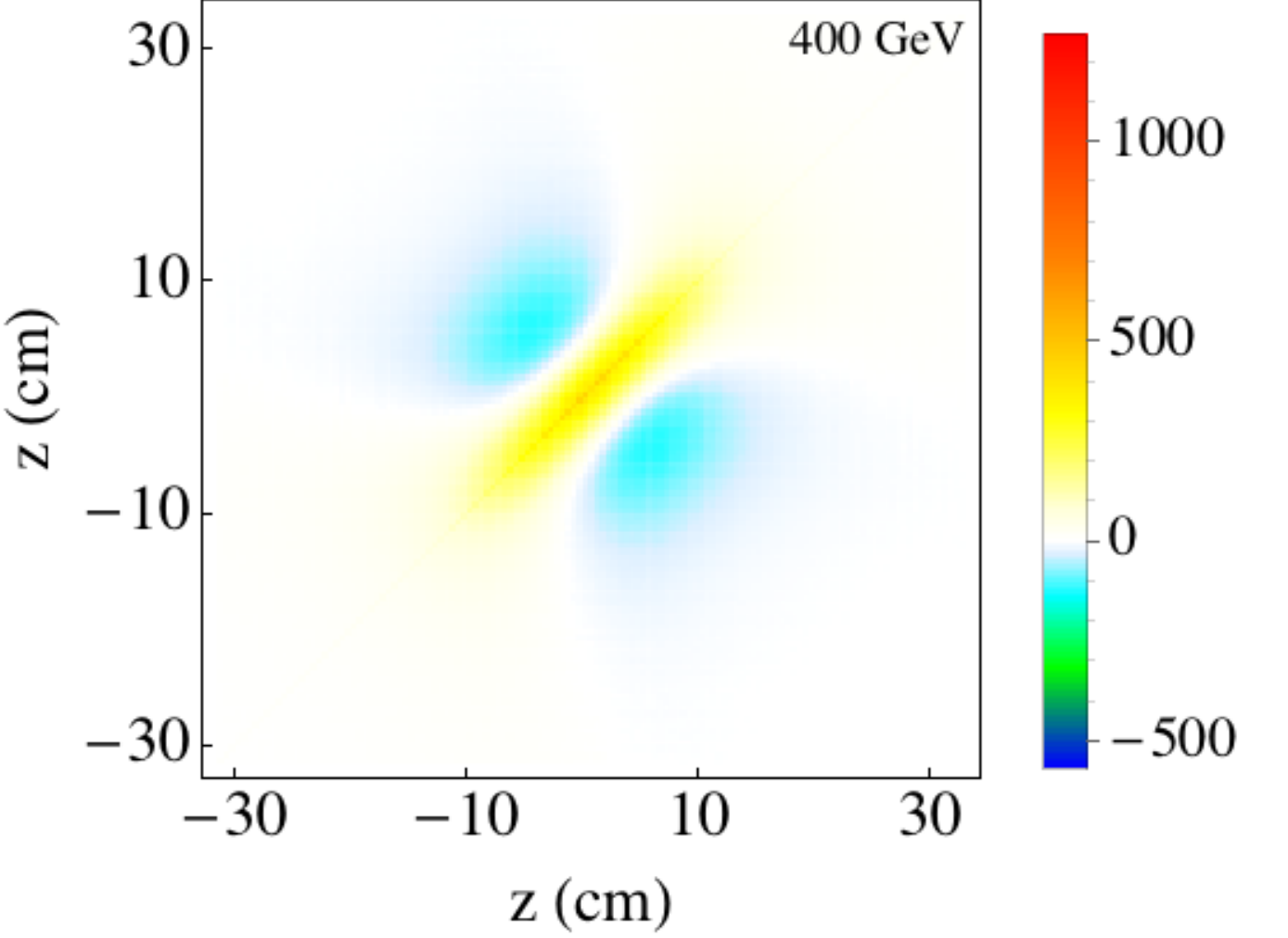}\\ \vspace{0.5cm}
\includegraphics[width=0.4\textwidth]{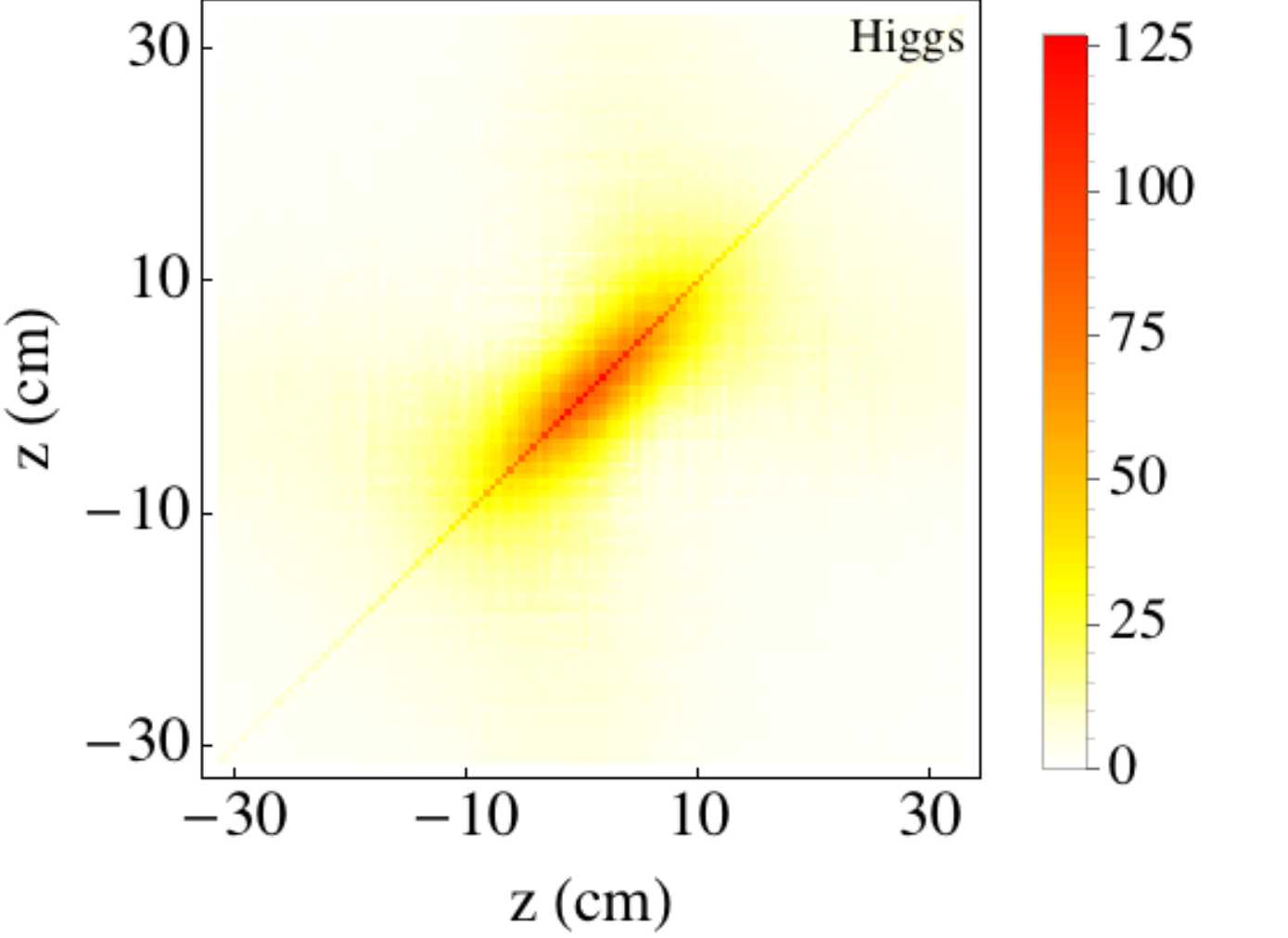}\hspace{1.5cm}
\includegraphics[width=0.4\textwidth]{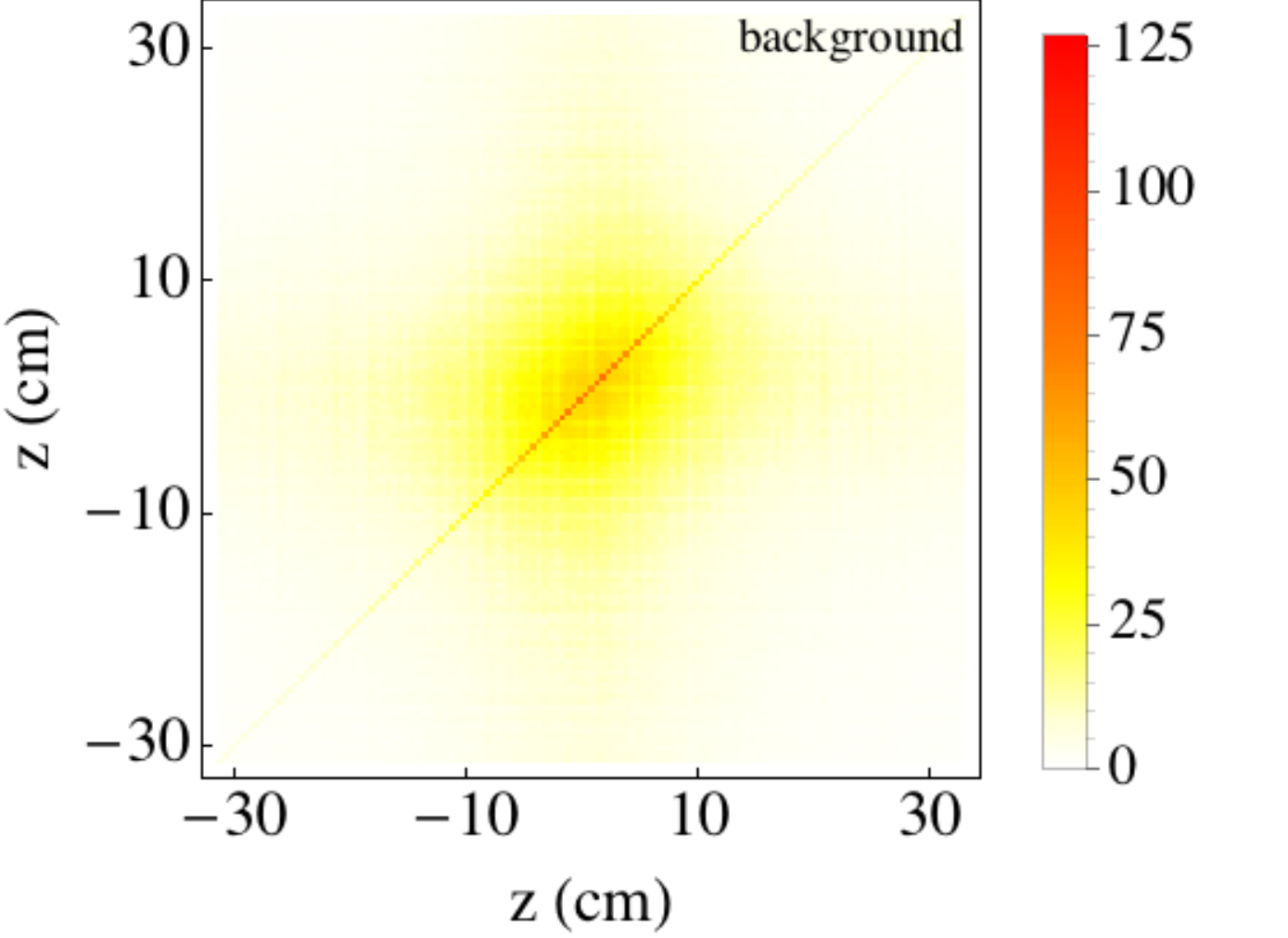}
\caption{Graphical representation of the covariance matrices for the dijet background and the three signal benchmarks. The scale in the bottom row is 1/10 of that in the top row.
\label{fig:covariance}}
\end{figure}

\begin{figure}[p]\centering
\includegraphics[width=0.4\textwidth]{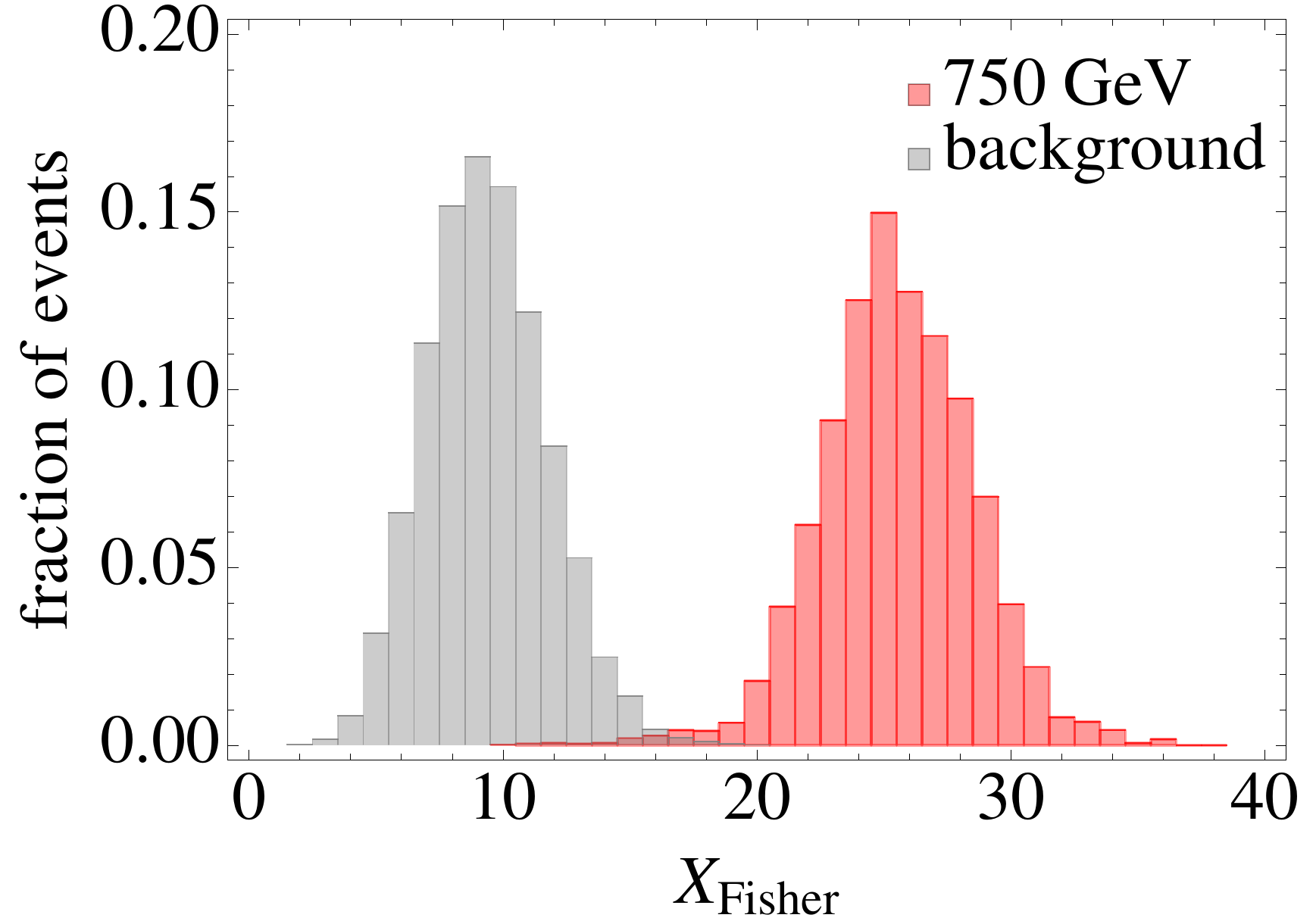}\hspace{1.5cm}
\includegraphics[width=0.4\textwidth]{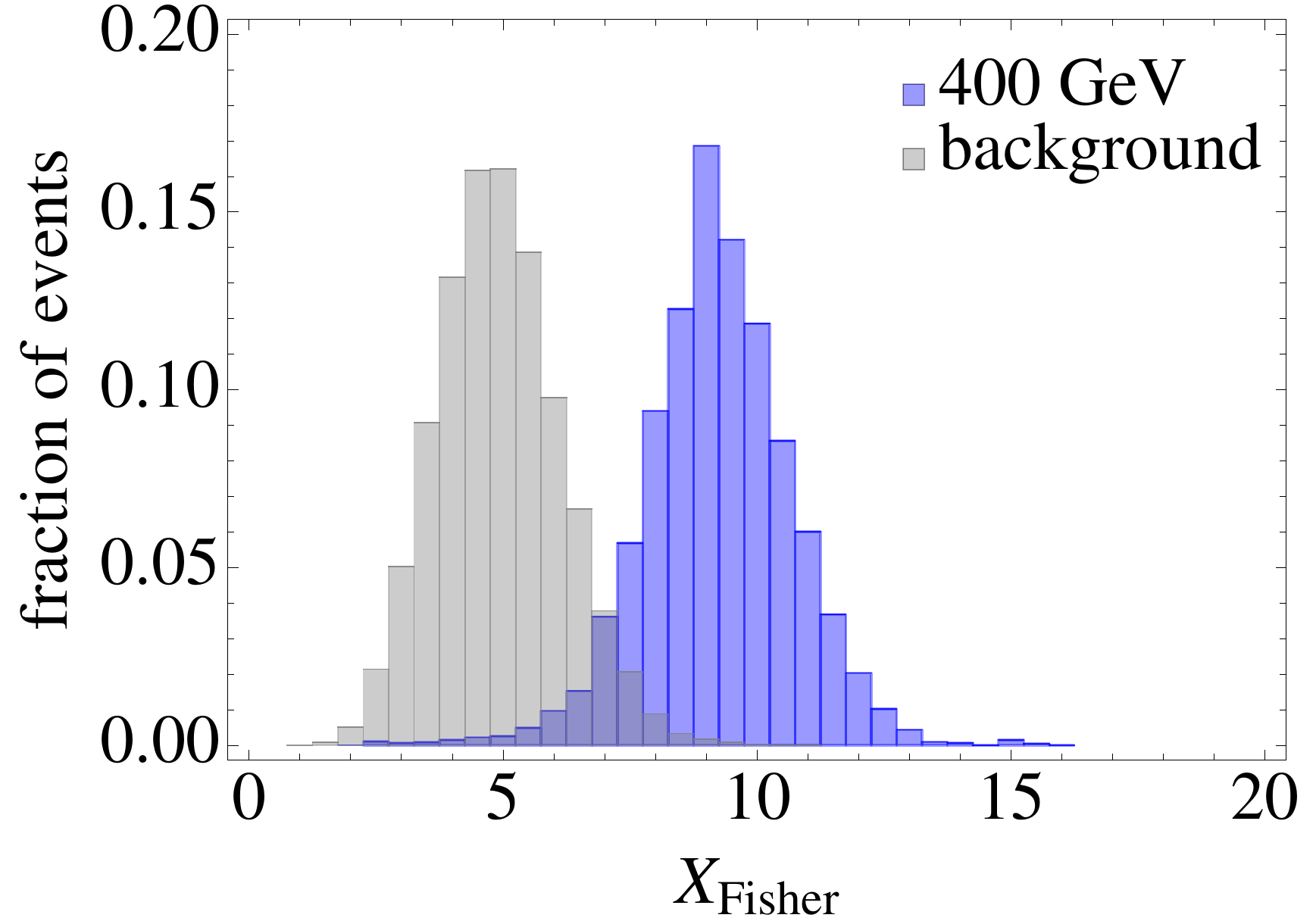}
\includegraphics[width=0.4\textwidth]{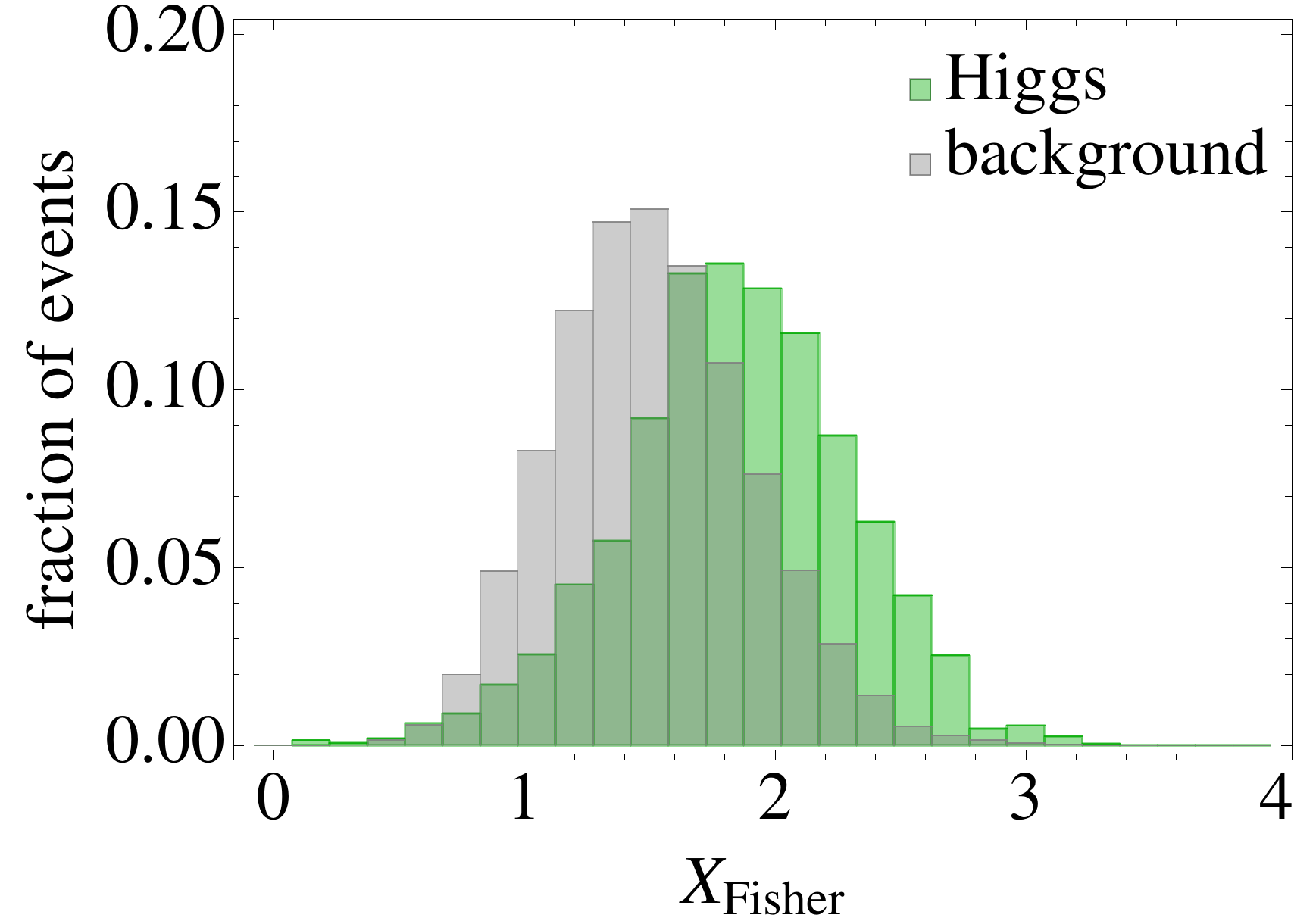}
\caption{Fisher discriminant for signal and background. \label{fig:fisher}}
\end{figure}

For our first variable, we choose to bin the data in $z$, with uniform bin width of $1$~cm, and compute the corresponding vector of expectation values $\vec 
\mu_{\bomb}$ ($\vec \mu_B$) for the number of hits in signal (background), both including pile-up. 
 We further compute the covariance matrices $\Sigma_{\bomb}$ and $\Sigma_B$ for all samples. Density maps of these covariance matrices are shown in Fig.~\ref{fig:covariance}. 
In particular, the signal exhibits large bin (anti)correlations, while the background is comparatively uncorrelated. 
Both populations are then separated by Fisher's linear discriminant, which is constructed as
\begin{equation}
X_{\text{fisher}}=(\vec \mu_{\bomb}-\vec \mu_B)\cdot(\Sigma_{\bomb}+\Sigma_{B})^{-1}\cdot\vec x
\end{equation}
where $\vec x$ is the data vector of a given event. The resulting distributions are shown in Figure~\ref{fig:fisher} for all three benchmarks. In Table~\ref{tab:HLTefficiency} 
we show the corresponding signal efficiencies for $10^{-3}$ and $10^{-4}$ background rejection rates. 

\begin{table}[b]
\renewcommand*{\arraystretch}{1.1}
\newcolumntype{C}{ >{\centering\arraybackslash} p{2cm} <{}}
\newcolumntype{D}{ >{\centering\arraybackslash} p{1.5cm} <{}}
\begin{tabular}{|D|D|D|D|D|D|D|D|D|}
\hline
\multirow{2}{*}{}	&\multicolumn{4}{c|}{Fisher}  	& \multicolumn{4}{c|}{Maximum Bin} \\\cline{2-9}
&\multicolumn{2}{c|}{BG rejection: $10^{-3}$}&\multicolumn{2}{c|}{BG rejection: $10^{-4}$}&\multicolumn{2}{c|}{BG rejection: $10^{-3}$}&\multicolumn{2}{c|}{BG rejection: $10^{-4}$}\\\cline{2-9}

	& $X^{\text{thres.}}_{\text{fisher}}$ & $\epsilon$ & $X^{\text{thres.}}_{\text{fisher}}$ & $\epsilon$ & $X^{\text{thres.}}_{\text{max}}$ & $\epsilon$ & $X^{\text{thres.}}
_{\text{max}}$ & $\epsilon$ \\ \hline
			
$750$ GeV	&18.4	&0.99	&20.25	&0.97	&1.99	&1.00	&2.19	&1.00\\
$400$ GeV	&9.63	&0.44	&10.41	&0.23	&1.99	&0.93	&2.19	&0.90\\
Higgs		&2.99	&0.009	&3.29	&0.0006	&1.99	&0.063	&2.19	&0.027\\\hline
\end{tabular}
\caption{Thresholds ($X^{\text{thres.}}_{\text{fisher}}$ and $X^{\text{thres.}}_{\text{max}}$) and corresponding signal efficiencies ($\epsilon$) for Fisher and Maximum Bin discriminators, for background rejection rates of $10^{-3}$ and $10^{-4}$. }
\label{tab:HLTefficiency}
\end{table}

\subsubsection{Maximum bin discriminant}

\begin{figure}[t]
\includegraphics[width=0.45\textwidth]{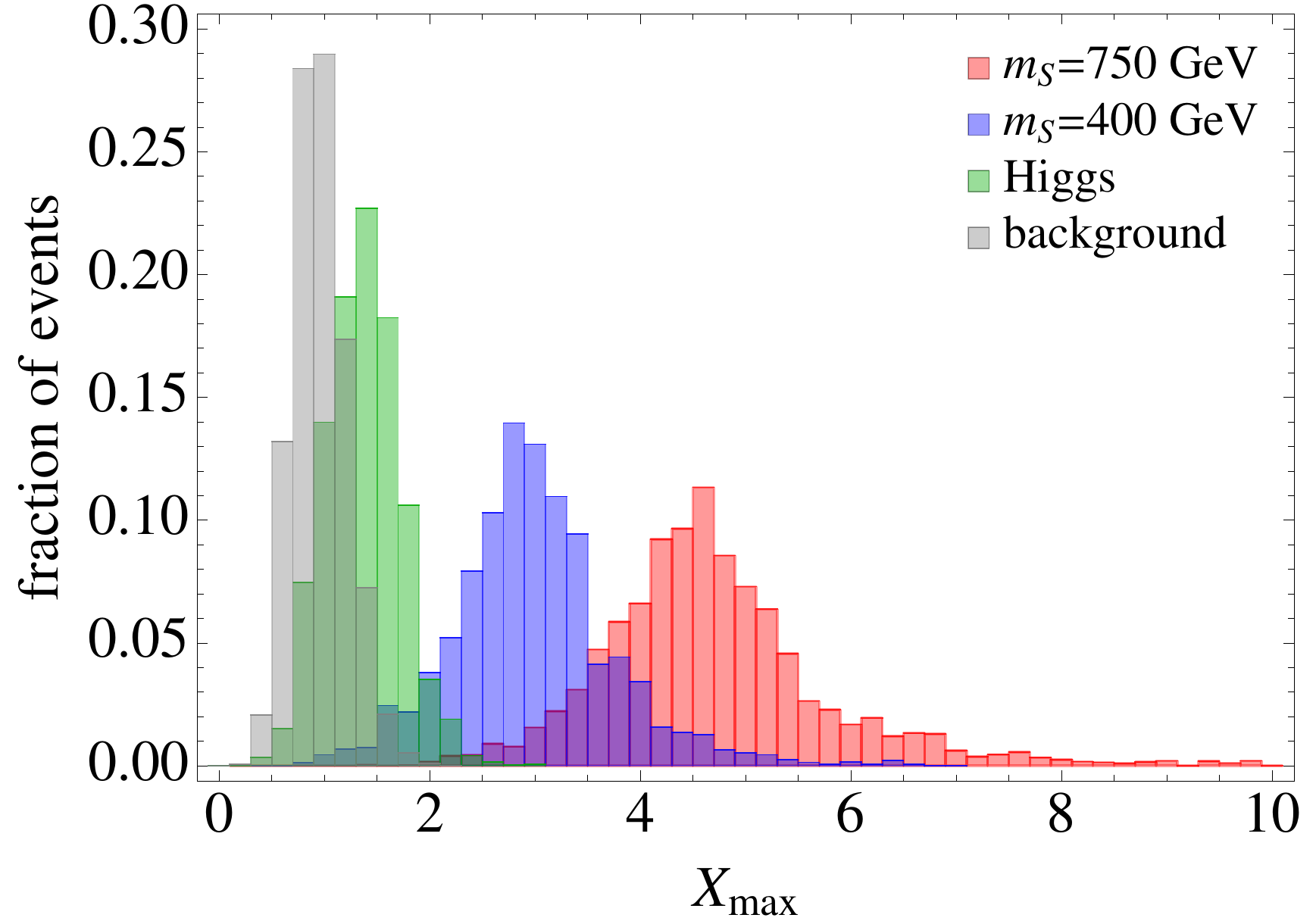}
\caption{$X_{\text{max}}$ discriminant. \label{fig:Xmax}}
\end{figure}

A disadvantage of the Fisher approach is that the $X_{\text{fisher}}$ threshold for a particular background rejection is dependent on the choice of signal benchmark. Our second variable 
uses information from the bins with the highest 
$S/B$ ratio, and has the advantage that it is independent of the signal properties. For each event we identify a 6 cm window around the bin with the highest number of hits. For this window we compute
\begin{equation}
	\label{eq:var2}
	X_{\text{max}}=\frac{\text{observed hit frequency}}{\text{expected background hit frequency}}\,.
\end{equation}
The resulting distributions are plotted in Fig.~\ref{fig:Xmax}. By construction, the background is centered around $X_{\text{max}}
\approx 1$, while the signal samples are shifted to higher values of $X_{\text{max}}$, such that it is possible separate signal and 
background in a model independent way by cutting on $X_{\text{max}}$.

The signal efficiencies for this variable are also shown in Table~\ref{tab:HLTefficiency}, for the two different background rejection levels. We find 
that the model independent variable somewhat outperforms the Fisher discriminant, and is nearly fully efficient for the 750 GeV and 
400 GeV benchmarks. For the Higgs portal benchmark we find a signal efficiency of roughly 10\%. This might be 
further improved, for instance, by weighting the 
likelihood that the background produces a maximum in the bin where the maximum was observed, rather than weighting by the expected amount of background, as in Eq.~\eqref{eq:var2}.
One could also include the information from the non-trivial $\phi$ dependence in Fig.~\ref{fig:trackereventphi}. Since our work is merely intended as a proof of concept, we do 
not attempt such further optimization here.

\begin{figure}[t]\centering
\includegraphics[width=0.3\textwidth]{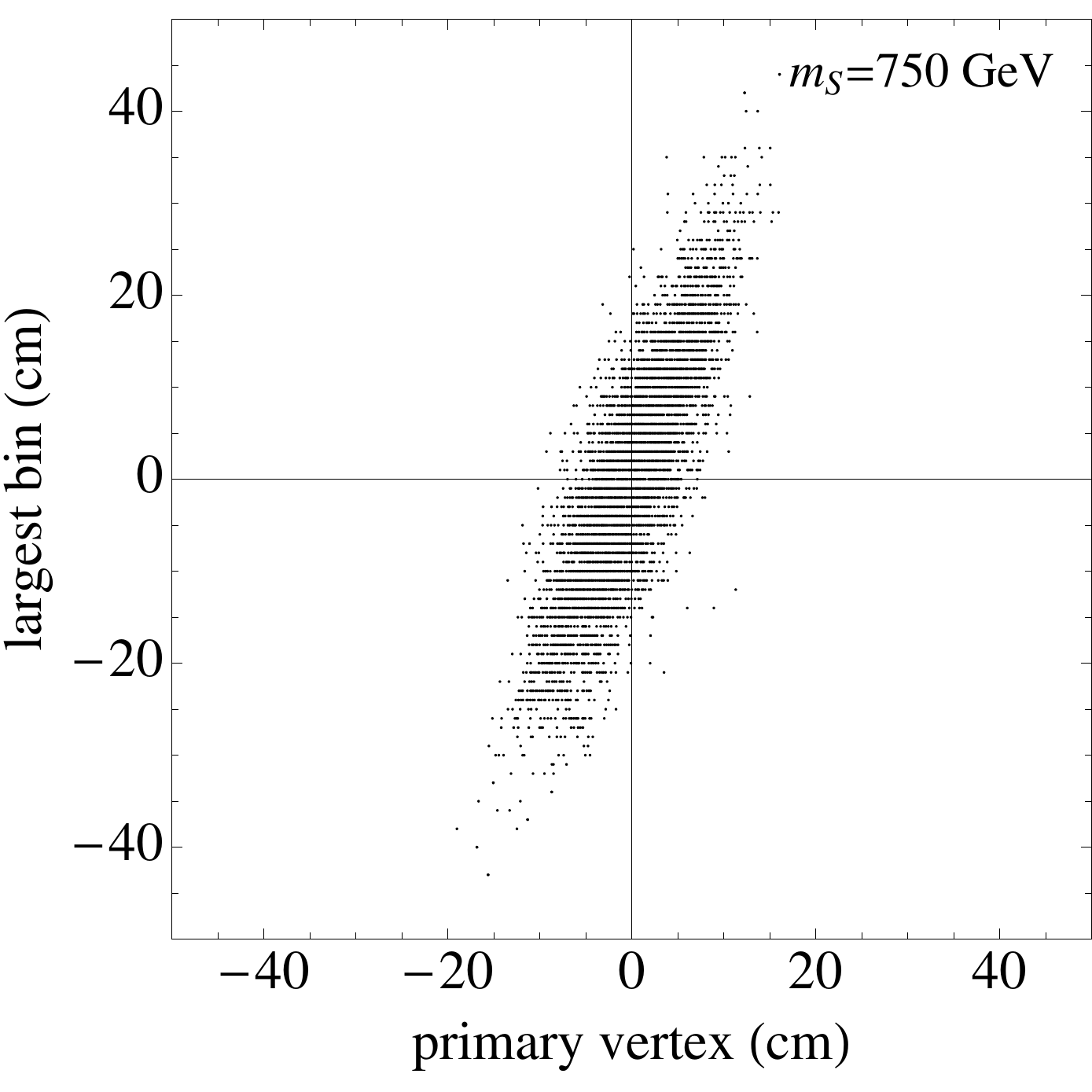}\hspace{2.5cm}
\includegraphics[width=0.3\textwidth]{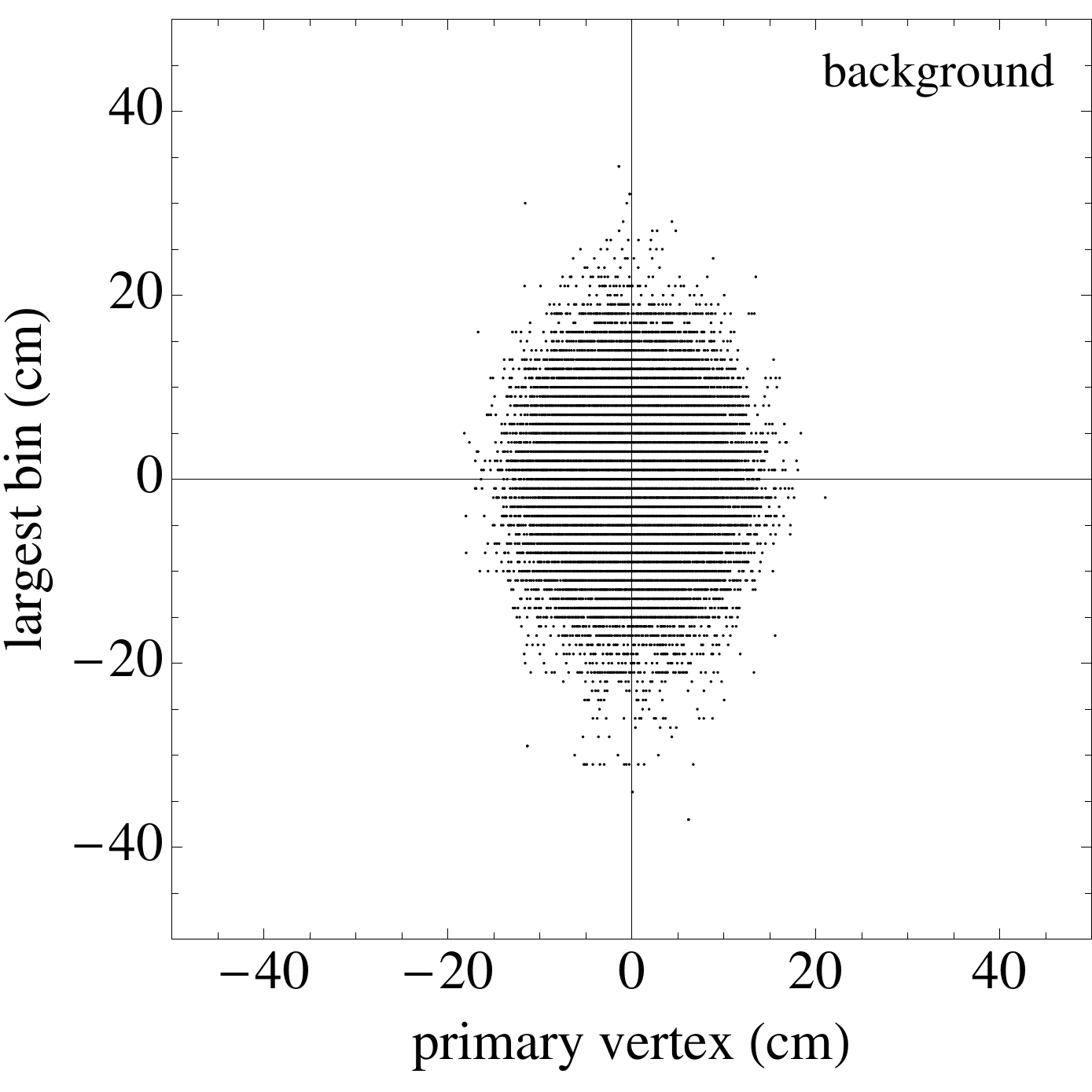}
\caption{Location of the truth-level primary vertex vs the location of the bin the largest number of hits. \label{fig:primaryvertex}}
\end{figure}

Further discrimination power may also be achievable if the primary vertex can be identified at the HLT level. In particular, for signal 
events one expects that the location of the bin with the largest number of hits is strongly correlated with the location of the primary 
vertex. Since the L1 trigger introduces a bias towards events with a hard ISR jet, we expect that the identification efficiency of the primary vertex should be relatively good, despite 
the softness of the tracks originating from the soft bomb itself.
 For the background on the other hand, the most populated bin is usually determined by fluctuations in the pile-up component. This means that its location 
is only weakly correlated with that of the primary vertex. This effect is shown in Fig.~\ref{fig:primaryvertex} for the 750 GeV benchmark compared to the background. 
Note that longitudinal boosts and the non-zero distance of the IBL from the interaction point smear out the distribution of the largest bin, compared to that of the primary vertex.
By themselves the distributions in Fig.~\ref{fig:primaryvertex} are less discriminating than the $X_{\text{fisher}}$ and $X_{\text{max}}$ 
variables. 
However, in a realistic analysis they may be complementary, since neither $X_{\text{fisher}}$ nor $X_{\text{max}}$ incorporated information regarding the location of the primary vertex.

\subsubsection{Muon HLT trigger\label{sec:muontrigger}}

\begin{table}[t]
\renewcommand*{\arraystretch}{1.1}
\newcolumntype{C}{ >{\centering\arraybackslash} p{1.2cm} <{}}
\newcolumntype{D}{ >{\centering\arraybackslash} p{2.75cm} <{}}
\newcolumntype{E}{ >{\centering\arraybackslash} p{2.5cm} <{}}
\begin{tabular}{|c|c|EC|CC|c|}
	\hline
Channel 		& $\sigma_{pp\rightarrow h+X }$~(pb) 	& L1 	&  HLT	& $\epsilon_{\mathrm{L1}}$ & $\epsilon_{\mathrm{HLT}}$ & $\sigma_{pp\rightarrow h+X }\times \epsilon_{\mathrm{L1+HLT}}$ (pb)\\
	\hline
GF/VBF/VH	& 50										& jet/$\MET$/$\mu$/$\gamma$/e & IBL 		& 0.071 & 0.06 & 0.21 \\
VH 			& 2.2										& $\mu$ & $\mu$						& 0.073 & 0.83  & 0.14 \\
	\hline
\end{tabular}
\caption{Comparison of IBL trigger path with the associated muon trigger path, assuming the $X_{max}$ variable with a level of background rejection of $10^{-3}$. \label{tab:assossummary}}
\end{table}

The combined L1 and HLT efficiencies for the Higgs portal benchmark using our IBL analysis are sufficiently low that the VH production channel using a muon trigger at HLT may be competitive, 
despite the lower cross section and the low branching ratio of the associated vector boson to muons. Triggerable muons at HLT are required to be isolated from other tracks with 
an isolation cone of outer (inner) radius $R=0.4$ ($R = 0.01$), and are required to be isolated from nearby jets with an isolation cone of $R = 0.4$. At the HLT-level, turn-on efficiency curves 
for the muon trigger are well-approximated by a step function at $p_T > 25$~GeV. In Table~\ref{tab:assossummary} we compare our belt of fire trigger on the IBL with this standard isolated 
muon trigger. For the IBL trigger path we combine the GF, VBF and VH production channels using the combined L1 trigger described in Sec.~\ref{sec:L1}. On the one hand, we find that both 
triggers paths are competitive, with our IBL trigger path slightly outperforming the single muon trigger. The single muon trigger, on the other hand, has the further advantage of being 
much simpler to implement, which should result in lower systematic uncertainties.

\section{Summary and Outlook}\label{sec:conclusions}

We presented a triggering strategy for events with an anomalously large multiplicity of soft tracks, which can occur in strongly-coupled hidden valley models that exhibit long, efficient 
showering windows. The central idea is to study hits rather than tracks at the HLT level, which could save a great deal of bandwidth and computation time. Specifically, we propose to search for a belt of fire: A local overdensity of hits on the innermost layers of the tracker. This approach provides excellent discriminating power for medium and high mass portals, to the extent that the HLT can be nearly 100\% efficient. 
We seed this new HLT trigger using the existing L1 jet or $\MET$ triggers by demanding a moderate amount of ISR. For the special case that the Higgs is the portal into the hidden valley, 
we compare the gluon fusion, vector boson and associated production modes, and include muon and electromagnetic L1 trigger paths.  The combined efficiency from our trigger is comparable to 
that which may be achieved by triggering on a hard muon from an associated vector boson. The combined L1+HLT trigger efficiencies for our benchmarks are summarized in Tab.~\ref{tab:summary}, 
where we also include an estimate of the sensitivity, requiring 3 events in 300 $\mathrm{fb}^{-1}$ of data. 

We have examined only a limited set of benchmarks in our analysis of this triggering strategy. However, to the extent that our benchmarks all produce a large number of soft SM states near 
the threshold of the detector reach, our strategy will be effective for any model that produces large numbers of soft but charged SM states. Although we only considered leptonic final states 
with no invisible particles in the final state, our analysis does not critically depend on this, so long as a sizable number of charged particles reach the IBL. It is also possible to consider 
the case that a sizable fraction of the hidden sector states are stable on collider time scales. This inevitably reduces the efficiency of our HLT strategy, to be likely more than offset 
by a more efficient L1 $\MET$ trigger. We leave a detailed exploration of these scenarios for future work.

\begin{table}[t]
\renewcommand*{\arraystretch}{1.1}
\newcolumntype{C}{ >{\centering\arraybackslash} p{2.9cm} <{}}
\newcolumntype{D}{ >{\centering\arraybackslash} p{1.5cm} <{}}
\begin{tabular}{|c|C|C|C|C|}\hline
&$750$ GeV& $400$ GeV&Higgs (all)& Higgs (VH)\\\hline
$\epsilon_{\mathrm{L1}}$&0.22&0.14&0.071&0.073\\
$\epsilon_{\mathrm{HLT}}$&1.00&0.93&0.06&0.83\\
$\epsilon_{\mathrm{L1+HLT}}$&0.22&0.13&0.004&0.06\\\hline
 $\sigma_{pp\rightarrow h+X }\times \epsilon_{\mathrm{L1+HLT}}$ (pb)&/&/&0.21 pb &0.14 pb\\\hline 
sensitivity (300 $\mathrm{fb}^{-1}$) & 0.045~fb & 0.078~fb & \resizebox{.99\hsize}{!}{$\mathrm{BR}_{h\rightarrow{\cal B}} < 4.8\times 10^{-5}$ }&  \resizebox{.99\hsize}{!}{$\mathrm{BR}_{h\rightarrow{\cal B}} < 7.1\times 10^{-5}$} \\\hline
\end{tabular}
\caption{Trigger efficiencies and estimated sensitivity for all benchmarks, assuming $10^{-3}$ level background rejection at the HLT. The first three columns contain the efficiencies for the belt of fire trigger. For the Higgs benchmark the GF, VBF and VH production modes are summed together. The last column presents the traditional single muon trigger path for VH production, as described 
in Sec.~\ref{sec:muontrigger}.\label{tab:summary}}
\end{table}

Another avenue is to consider the case that the hidden sector states decay displaced inside the inner detector, requiring consideration of more outlying layers of the tracker in addition 
to the IBL. Applying our proposed trigger on some of the outer tracker layers may still work, although with a reduced efficiency because of the more diffuse distribution of the signal hits. 
However this scenario may also provide new handle in the ratios of the number of hits between the various layers.

 Given that LHCb will eventually operate fully in the trigger-less mode, it will have unique sensitivity to soft signatures of new physics, as demonstrated in~\cite{Ilten:2015hya,Ilten:2016tkc}.  
 It may also be possible to search for soft, displaced jets with the VELO system of LHCb detector~\cite{yueinprogress}. For the soft bomb scenario in particular, it should be possible to apply 
 a version of our proposal as a fully online analysis at LHCb. Since LHCb essentially bypasses the L1 trigger, this may be particularly useful for soft bombs near the low mass benchmark, for which 
 we found the L1 efficiency at ATLAS to be comparatively low.  Even with its lower luminosity, LHCb may thus favorably compete with the lower efficiency but higher luminosity at ATLAS and CMS.

\textbf{Acknowledgments}\\
We thank Brian Batell, Jean-Paul Chou, Kyle Cranmer, Maurice Garcia-Sciveres, Andrew Haas, Ian Hinchcliffe, Markus Klute, Greg Landsberg, Henry Lubatti, Matt Strassler, Jesse Thaler,  
Yue Zhao  and Kathryn Zurek for useful discussions. This work was supported in part by the Department of Energy (DoE) under contract DE-AC02-05CH11231, and by the National Science Foundation (NSF) under 
grant No. PHY-1002399. This research used resources of the National Energy Research Scientific Computing Center, which is supported by the Office of Science of the DoE under 
Contract No.~DE-AC02-05CH11231. SK, MP and DR thank the Aspen Center for Physics, supported by the NSF under grant No.~PHY-1066293, for hospitality while parts of this work were completed. SK and MP also thank the Gallileo Galilei Institute for Theoretical Physics where part of this work was completed. DR acknowledges support from the University of Cincinnati. 

\appendix

\section{Simulation of the detector response\label{app:detecsim}}
In this section we describe our simulation of the response of the ATLAS inner detector, as well as the various approximations that were 
used. This simulation is designed to estimate the distribution of hits on IBL, as well as the energy distribution in the electromagnetic calorimeter. The 
latter is required in order to estimate the efficiency of the L1 trigger. Given that many tracks are soft, energy loss through ionization and 
bremsstrahlung in the various layers of the inner detector can significantly modify the trajectory of the particles, which in turn modifies 
the distribution of the hits on the IBL. The photons emitted through bremsstrahlung of the electrons and positrons moreover produce a 
more or less uniform haze of energy in electromagnetic calorimeter, which could affect the L1 trigger efficiencies. 

The components included in our simulation are
\begin{itemize}
\setlength\itemsep{-0.4em}
\item beam pipe
\item IBL, one layer in the barrel
\item pixel detector, consisting out of 3 layers in the barrel 
\item Silicon microstrip tracker (SCT), consisting out of 3 layers in the barrel and 2 $\times$ 9 discs in the endcaps
\item Service layer
\item Transition Radiation Tracker (TRT), consisting out of a barrel and endcap
\item Electromagnetic Calorimeter (ECAL), consisting out of a barrel and endcap.
\end{itemize}
We take the resolution and geometry of each component from the ATLAS technical design report~\cite{Airapetian:391176}. To establish the interaction points, all components are taken to be 
infinitely thin and located at their mean position, except for the TRT barrel, for which we divided the radial dimension into 4 equal 
segments, each are taken to be infinitely thin. We propagate all charged particles in the magnetic field, which for ATLAS we take to 
be 2 Tesla and uniform throughout the detector volume. We stop their trajectories when one of the following conditions is met:
\begin{enumerate}
\setlength\itemsep{-0.4em}
\item They reach the calorimeter wheel or barrel.
\item Their kinetic energy drops below 10\% of their mass.
\item Their time-of-flight exceeds 25 ns.
\item They escape down the forward regions not covered by the calorimeters.
\end{enumerate}
Neutral particles are always propagated straight to the calorimeter barrel or wheels without energy loss in the inner tracker. For soft 
photons this is not necessarily a good approximation, however it is nevertheless likely to be conservative, since it corresponds to a slight
overestimation of the uniform haze in the electromagnetic calorimeter.

Energy loss per unit distance through ionization is modeled with the Bethe-Bloch equation~\cite{Olive:2016xmw}, considering the most probable value of
the energy loss (a better approximation for thin layers) and neglecting straggling. For each component we compute total energy loss along the path travelled through the material, 
accounting for the incident angle of the particle. For electrons and positrons we also compute the energy loss through bremsstrahlung 
\cite{Olive:2016xmw}, where we obtained radiation length of the various components from~\cite{Andreazza:2009zz}. All bremsstrahlung energy is 
assumed to go into a single photon, which is emitted tangentially to the direction of motion.  Bremsstrahlung photons are then propagated 
to the electromagnetic calorimeter without further energy loss.

Whenever a particle crosses the IBL, we count this as a single hit. In reality particles with small incident angle may cross multiple pixels, 
something which ATLAS accounts for by clustering neighboring pixels. We estimate that the likelihood of two different soft bomb 
decay products lighting up two neighboring pixels is less than $1\%$, and therefore identify `hits' in our simulation with `clusters' in the 
ATLAS detector.  We validate our estimates against the publicly available data as follows: From public plots~\cite{ATLASpixelpublic} we 
find that the occupancy of the IBL and pixel barrel layer 0 are comparable in a high pile-up environment. Extrapolating to an 
average of $50$ pile-up interactions, we expect roughly 2400 clusters in the IBL on average. This is roughly 30\% higher that than what we 
obtain with our simplified simulation. The difference can largely be attributed to the fact that \texttt{Pythia 8} somewhat underestimates the number 
of the soft tracks in the minimum bias events, as further indicated by the results of~\cite{Aaboud:2016itf}. Given that our choice for the pile-up conditions is roughly a factor of two higher than 
current running conditions, we consider this error to be acceptable.

\label{app:CM}

\bibliography{bombs}

\end{document}